\documentclass[aps,prb,twocolumn,showpacs,amsmath,amssymb,superscriptaddress]{revtex4-2}
\usepackage{graphicx}
\usepackage{CJK}
\usepackage{color}
\usepackage[colorlinks,bookmarks=false,citecolor=blue,linkcolor=red,urlcolor=blue]{hyperref}
\usepackage{multirow}
\usepackage{ulem}
\usepackage{tabularx}
\usepackage[nounderscore]{syntax}
\graphicspath{{figs/}}
\usepackage{subfigure}
\usepackage{float}

\newcommand{\be}{\begin{equation}}
\newcommand{\ee}{\end{equation}}

\begin{document}
\title{Absence of logarithmic and algebraic scaling entanglement phases due to skin effect}

\author{Xu Feng}
\altaffiliation{The two authors contributed equally to this work.} 
\affiliation{Department of Physics, Beijing Normal University, Beijing 100875, China}

\author{Shuo Liu}
\altaffiliation{The two authors contributed equally to this work.} 
\affiliation{Institute for Advanced Study, Tsinghua University, Beijing 100084, China}

\author{Shu Chen}
\email{schen@iphy.ac.cn }
\affiliation{Beijing National Laboratory for Condensed Matter Physics, Institute
of Physics, Chinese Academy of Sciences, Beijing 100190, China}
\affiliation{School of Physical Sciences, University of Chinese Academy of Sciences,
Beijing 100049, China }
\affiliation{Yangtze River Delta Physics Research Center, Liyang, Jiangsu 213300,
China }

\author{Wenan Guo}
\email{waguo@bnu.edu.cn}
\affiliation{Department of Physics, Beijing Normal University, Beijing 100875, China}
\affiliation{Beijing Computational Science Research Center, Beijing 100193, China}

\date{\today}
\begin{abstract}
Measurement-induced phase transition in the presence of competition between projective measurement and random unitary evolution has attracted
increasing attention due to the rich phenomenology of entanglement structures. However, in open quantum systems with free fermions, a generalized
measurement with conditional feedback can induce skin effect and render the system short-range entangled without any entanglement transition, 
meaning the system always remains in the ``area law'' entanglement phase. In this work, we demonstrate that the power-law long-range hopping does 
not alter the absence of entanglement transition brought on by the measurement-induced skin effect for systems with open boundary conditions. 
In addition, for the finite-size systems, we discover an algebraic scaling $S(L, L/4)\sim L^{3/2-p}$ when the power-law exponent $p$ of long-range 
hopping is relatively small. For systems with periodic boundary conditions, we find that the measurement-induced skin effect disappears 
and observe entanglement phase transitions among ``algebraic law'', ``logarithmic law'', and ``area law'' phases. 

\end{abstract}
\maketitle


\section{Introduction}
\label{intro}

The rich phenomenology of entanglement structures has sparked increased interest in measurement-induced phase transition (MIPT) \cite{Nahum, FisherPRB1, FishePRB2, unitaryproj, PhysRevLett.126.060501, liuUniversalKPZScaling2022, PhysRevX.12.011045, PRXQuantum.2.040319, MIPT_Clifford_PRB19_Szyniszewski, MIPT_Clifford_PRL20_Qi, NonlocalMIPT_PhysRevX, MIPT_Clifford_PRB20_Bao, MIPT_Clifford_PRB21_Fan, MIPT_Clifford_PRB21_Li, MIPT_Clifford_PRB21_Jian}.
A prototypical model exhibiting MIPT is the monitored quantum systems undergoing random unitary evolution interspersed by local projective measurements, in which the random unitary evolution tempts to induce large-scale entanglement while the measurements shatter quantum coherence 
and thereby suppress entanglement growth. Below a critical measurement rate $p_c$, the entanglement within the system obeys ``volume law''. 
Increasing the measurement rate above the critical rate, the system enters the ``area law'' entanglement phase \cite{Nahum, FisherPRB1, FishePRB2,unitaryproj}. 
Recently, MIPT has also been investigated in the monitored free fermion chain and quantum Ising chain \cite{fermionMIPT, DiehlMIPT, Isingchain, Isingchain2, Isingchain3, Isingchain4, EffTheoryMIPT,topofermions}. With increasing the rate of the continuous local number measurement, both 
systems undergo an entanglement transition from the ``sub-extensive law" phase into the ``area law" phase.

To further understand the nature of the measurement-induced phase transition, the effect of the long-range interaction between qudits (or long-range hopping between free fermions) needs to be considered and remains an essential open question, while the previous studies mainly focus on the monitored systems with local or all-to-all interaction (hopping). In principle, the long-range interaction (hopping) may change the universality class and even induce a richer entanglement phase diagram. For power-law long-range interacting hybrid quantum circuits, long-range interactions give rise to a continuum of non-conformal universality classes and induce a novel ``sub-volume law" phase \cite{longrangecircuit}. Analogously, for monitored 
power-law hopping free fermion model, long-range hopping induces an unconventional algebraic scaling phase when power-law hopping decay 
exponent $p\leq d/2+1$ where $d$ is spatial dimension \cite{longrangeMIPT}. If the power-law long-range interaction is also added, the critical exponent $p_c$ for the algebraic scaling phase will be increased to $d+1$ \cite{longrangeMIPT2}.

Besides the local measurements, other mechanisms such as non-Hermitian skin effect (NHSE) \cite{GBZ, photonic, WindingSkin,windingSkin2, Anatomyskinmodes,boundarymodes,non-bloch,auxiliaryGBZ,boundarysensitive, CNHSE,exceptional, EEntropyNonHerm} can also suppress entanglement 
growth and recent work has revealed that the entanglement of free fermion systems undergoing non-Hermitian evolution \cite{NHSEEE}, i.e., the 
post-selection quantum trajectory without quantum jumps obeys ``area law''. Moreover, the suppression effect for entanglement due to skin effect 
has also been investigated in the presence of quantum jumps, in which the monitored free fermion model includes nearest-neighbor hopping and
generalized measurements with conditional feedback. When the measurement rate is non-zero, particles tend to accumulate at one specific edge. The 
measurement-induced skin effect causes most particles to freeze, rendering the system short-range entangled. Therefore, the entanglement averaged 
over the full trajectories also obeys ``area law''. It is worth noting that the skin effect relies on open boundary conditions (OBCs).

An interesting and vital question is, what are the entanglement behaviors for the free fermion systems with long-range hopping and 
measurement-induced skin effect? Theoretically, long-range hopping can spread quantum information to particles at any distance, making it possible 
to overcome localization brought on by the skin effect and sustain large-scale entanglement. 

To answer this question, we investigate the entanglement behaviors of a monitored free fermion model with power-law long-range hopping and 
conditional generalized measurement in this work.
When periodic boundary conditions (PBCs) are adopted, we find that the skin effect disappears. We obtain an entanglement phase diagram (see Fig. \ref{phasediagram}(a)) including ``area", ``logarithmic" and ``algebraic law" phases via tuning the power-law exponent $p$ and the 
generalized measurement rate $\gamma$. Our numerical results are consistent with the previous study in which the generalized measurement is 
replaced by measurement for local density \cite{longrangeMIPT}. While with OBCs, we obtain convincing numerical results in support that the 
skin effect survives under the power-law long-range hopping. Despite the system ultimately entering the ``area law'' entanglement phase in 
the thermodynamics limit (see Fig. \ref{phasediagram}(b)), we show that finite-size effects can induce different entanglement behaviors via tuning 
the parameters. When $p$ is large, we find that the entanglement behaviors are similar to the case with nearest hoppings ($p \rightarrow \infty$), 
and the entanglement decays to zero fastly with the increasing of the system size $L$ \cite{MISE}. Nevertheless, when $p$ is small, we find 
an algebraic scaling behavior $S(L, L/4) \sim L^{\frac{3}{2}-p}$ for finite-size system. 


The paper is organized as follows. In Sec. \ref{Sec:II}, we introduce the model studied in this work:  a monitored free fermion model with 
power-law long-range hopping and generalized measurement and the observables that quantify the entanglement and the skin effect. 
Next, in Sec. \ref{sec:longrange}, we perform numerical simulations using the quantum jump approach and show the numerical results of 
different Hamiltonian parameters and boundary conditions.
Then we analyze the entanglement entropy, classical entropy, and local density distributions of steady states to show the competition between 
long-range hopping and the measurement-induced skin effect, thus inferring the phase diagram. Moreover, in Sec. \ref{sec:nofeedback}, we 
investigate the case without conditional feedback, in which the numerical results are sensitive to time step $\delta t$ in quantum jump simulation
and observe a ``pseudo skin effect" when $\delta t$ is not small enough.
Finally, in Sec. \ref{sec:conclsn}, we give our conclusions and some outlooks. Technical details about numerical simulation are described in 
Appendix \ref{sec:appendixA}. In \ref{sec:appendixB}, we show more supportive numerical results.

\begin{figure}[t]
\centering
\centering
\includegraphics[height=3.0cm,width=8.5cm]{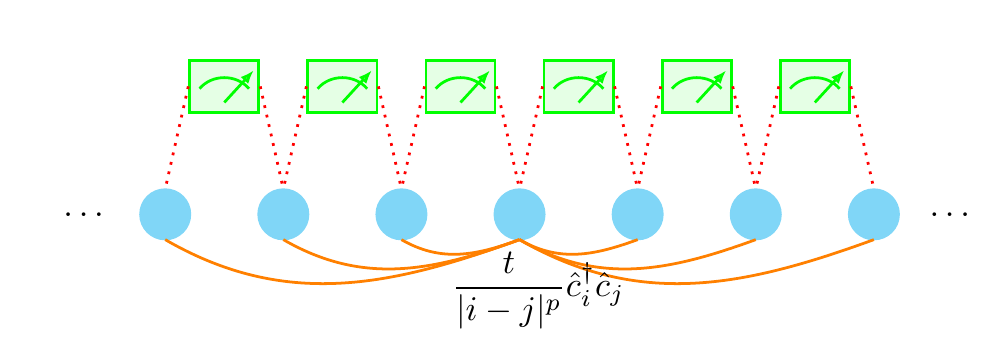}

\caption{The schematic diagram for the free fermion model with power-law long-range hopping and generalized continuous measurements. The generalized measurements (green block) act on every pair of neighboring sites. }
\label{model}
\end{figure}

\section{Model and Measurement Protocols}
\label{Sec:II}
We consider a one-dimensional (1D) free fermion model with power-law long-range hopping. The Hamiltonian is as follows:
\begin{equation}\label{eq1}
	\begin{split}
    \hat{H}_{0}=\sum_{i\neq j}\dfrac{t}{|i-j|^{p}} \hat{c}^{\dagger}_{i}\hat{c}_{j},
     \end{split}
	\end{equation}
where $\hat{c}_{i}$ and $\hat{c}_{i}^{\dagger}$ are annihilation and creation operators of the spinless fermion at site $i$, respectively, $t$ is the 
hopping strength, and $p$ is the exponent determining the hopping range. We set $t=1$ throughout the work. To avoid the singular fermion dispersion, 
we mainly focus on the case $p>1$ \cite{longrangeMIPT}.

Different from projective measurements in the hybrid quantum circuits, the continuous measurement process for the free fermion model, which is
an open quantum system, is a kind of weak measurement, and the monitoring dynamics is described by the stochastic Schrodinger equation (SSE) \cite{quantum_measurement_and_control,continuous_quantum_measurement,quantum_noise,quantumtraj,OQS},
\begin{equation}\label{eq2}
	\begin{split}
	d|\psi\rangle &= -i \hat{H}_{\text{eff}}|\psi\rangle\,dt + \sum_{\mu}[\dfrac{\hat{L}_{\mu}}{\sqrt{\langle \hat{L}^{\dagger}_{\mu}\hat{L}_{\mu\rangle}}}-1]|\psi\rangle\,dW_{\mu},
		\end{split}
\end{equation}
where $\hat{L}_{\mu}$ is the quantum jump operator modeling the conditional monitored observable, and each $dW_{\mu}$ is a discrete, independent
Poisson random variable $dW_{\mu}=0$ or $1$, with mean value $\overline{dW_{\mu}}=\gamma\langle\hat{L}_{\mu}^{\dagger}\hat{L}_{\mu}\rangle dt$ 
in which $\gamma$ is the monitoring rate. And the effective Hamiltonian is
\begin{eqnarray}
\hat{H}_{\text{eff}}=\hat{H}_{0}-i\dfrac{\gamma}{2}\sum_{\mu}\hat{L}_{\mu}^{\dagger}\hat{L}_{\mu}.
\end{eqnarray}

\subsection{Quantum jump operator}
Inspired by \cite{MISE}, we choose the quantum jump operator
\begin{equation}
	\begin{split}
	\hat{L}_{i} &= \hat{U}_{i}\hat{P}_{i}=\dfrac{1}{2}\,  e^{i\theta\hat{n}_{i+1}}\,  \hat{\xi}^{\dagger}_{i}\, \hat{\xi}_{i},
		\end{split}
	\end{equation}
where unitary feedback $\hat{U}_{i}=e^{i\theta \hat{n}_{i+1}}$ and measurement operator $\hat{P}_{i}=\dfrac{1}{2}\hat{\xi}^{\dagger}_{i}\hat{\xi}_{i}$ with $\hat{\xi}^{\dagger}_{i}=\hat{c}^{\dagger}_{i}-i\hat{c}^{\dagger}_{i+1}$. We focus on the case with $\theta=\pi$ firstly, which is more 
intuitive for measurement-induced skin effect and defer the discussions about the case with $\theta=0$, i.e., no conditional feedback, to Sec. \ref{sec:nofeedback}. 

Physically, $\hat{\xi}^{\dagger}_{i}$ can be regarded as the creation operator of a right-moving quasi-mode. Because  $\hat{\xi}_{i}^{\dagger}=\sum_{k}g(k)\hat{c}_{k}^{\dagger}$ and  $|g(k)|^2\propto (1-\sin k) $, consequently $\hat{\xi}_{i}^{\dagger}$ induces imbalance particle distributions in momentum space. Therefore, the current $J=\int_{-\pi}^{\pi}v_{k}n_{k}$, where $v_{k}$ for small monitoring rate $\gamma$ can be approximated as power-law hopping free fermion case $v_{k}=\partial_{k} E(k)=-2\sum_{m}m \sin (mk)/m^{p}$, is positive, which indicates the right moving quasi-mode. The conditional feedback $\hat{U}_{i}$ 
is for converting the right-moving quasi-mode $\hat{\xi}^{\dagger}_{i}=\hat{c}^{\dagger}_{i}-i\hat{c}^{\dagger}_{i+1}$ into left-moving quasi-mode $\hat{\xi}^{\prime \dagger}_{i}=\hat{c}^{\dagger}_{i}+i\hat{c}^{\dagger}_{i+1}$ $(|g(k)|^2\propto (1+\sin k))$. Past research shows that for the nearest hopping case, the conditional feedback is indispensable for measurement-induced skin effect; otherwise, the right-moving quasi-mode created by quantum jumps is canceled by the left-moving quasi mode induced by $\hat{H}_{\text{eff}}$. In fact, there is huge freedom to choose $\theta$. However, the numerical results reveal that with the decrease of $\theta/\pi$, the fluctuation of the boundary between occupied and unoccupied regions gets extended until the measurement-induced skin effect can not be observed in the $\theta  \rightarrow 0$ limit. More details and discussions can be found in Ref. \cite{MISE}.

\begin{figure}[htb]
\centering
\centering
\includegraphics[height=5.0cm,width=7.0cm]{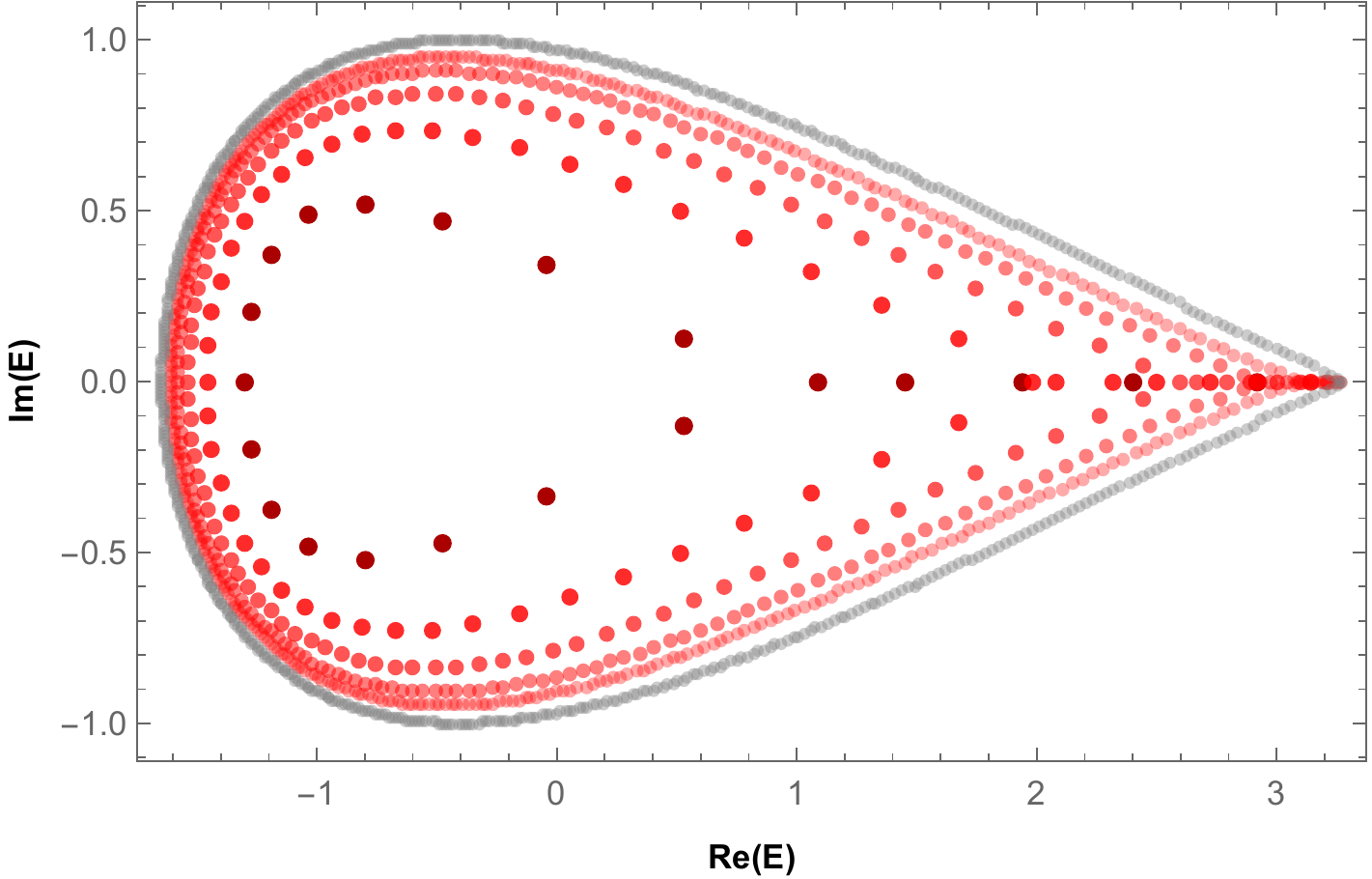}

\caption{Energy spectrum of $\hat{H}^{\prime}_\text{eff}$ with $p=2.0$, $\gamma=2.0$, and $L=20,  50, 100, 200, 400$ under OBC. Dark (light) color corresponds to small (large) system size $L$, and the OBC spectrum approaches the PBC spectrum (Gray) as $L$ increases.}
\label{H_eff}
\end{figure}

\subsection{The effective Hamiltonian $\hat{H}_{\text{eff}}$}

The effective non-Hermitian Hamiltonian $\hat{H}_\text{eff}$ is
 \begin{equation}
	\begin{split}
	\hat{H}_{\text{eff}} =& \sum_{i\neq j}\dfrac{t}{|i-j|^{p}}\hat{c}^{\dagger}_{i}\hat{c}_{j} \\ 
	&+\dfrac{\gamma}{4}\sum_{i}( \hat{c}^{\dagger}_{i}\hat{c}_{i+1}- \hat{c}^{\dagger}_{i+1}\hat{c}_{i}-i(\hat{n}_{i}+\hat{n}_{i+1})).
		\end{split}
	\end{equation}
Since overall dissipation only gives a shift to the spectrum, we will directly analyze $\hat{H}^{\prime}_\text{eff}$'s energy spectrum, where 
\begin{eqnarray}
\hat{H}^{\prime}_\text{eff}=\sum_{i\neq j}\dfrac{t}{|i-j|^{p}}\hat{c}^{\dagger}_{i}\hat{c}_{j}
	+\dfrac{\gamma}{4}\sum_{i}( \hat{c}^{\dagger}_{i}\hat{c}_{i+1}- \hat{c}^{\dagger}_{i+1}\hat{c}_{i}).
\end{eqnarray}
As shown in Fig. \ref{H_eff}, for small system size $L$, the OBC eigenstates are clearly distinct from PBC eigenstates. However, with the increase of $L$, the OBC energy spectrum approaches the PBC energy spectrum gradually. 

We now discuss $\hat{H}_\text{eff}^{\prime}$'s single-particle OBC eigenstates. In the $p \rightarrow \infty$ limit, it is exactly 
the Hatano-Nelson model \cite{HNmodel}. The magnitude of the single-particle OBC eigenstates exponentially decays with size-independent 
localization length. While for $p=0$, i.e., the all-to-all limit, there is almost no difference between PBC and OBC, and NHSE is absent. 

Beyond the single-particle case,
for the Hatano-Nelson model ($p=\infty$), the density imbalance between two half sides
grows linearly with the system size $L$, i.e., the skin effect still exists for many-body systems  \cite{manybodyNHSE}. However, when $p$ is finite, 
it's reasonable to anticipate that the density imbalance grows more slowly with $L$ or even saturates for large $L$, i.e., the skin effect may be suppressed or destroyed by long-range hopping. Worth to mention for any $\gamma>0$, skin modes induced by $\hat{H}^{\prime}_\text{eff}$ tend to localize at the left side.

The conditional feedback is necessary for measurement-induced skin effect when $p = \infty$ \cite{MISE}. Otherwise, the quasi-modes with opposite directions induced by the quantum jump and $\hat{H}_{\text{eff}}$ will probably cancel each other.
When the skin effect emerges, due to the Pauli exclusion principle, most of the particles will freeze, thus suppressing the quantum correlation developed by hopping and leading to ``area law" entanglement \cite{MISE}.
It is worth noting that the measurement-induced skin effect only can be revealed with OBC. To demonstrate the different entanglement behaviors caused by the boundary conditions, both cases with PBC and OBC are investigated in this work.

\subsection{Observables}
Firstly, to quantify the strength of the skin effect, we introduce 
\begin{equation}
	\begin{split}
\Delta n&= |N_\text{left}-N_\text{right}|/N_\text{tot},
	\end{split}
	\end{equation}
where $N_\text{left}$ ($N_\text{right}$) represents the number of particles in the left (right) half side, and $N_\text{tot}$ means the total particle number. 
For strong skin effect, particles almost accumulate at one side, thus $\Delta n \approx 1$. While if density distribution is uniform, $\Delta n \approx 0$.

Since both the $\hat{H}_{\text{eff}}$ and monitoring operator $\hat{L}_{\mu}$ are quadratic, the dynamical evolution preserves gaussianity with 
a Gaussian initial state. 
For the Gaussian state, the bipartite entanglement entropy with subsystem $A=[1, L/4]$ is given by \cite{EEntropyevolution,cauchyformula,Peschel_2009,Peschel_2004,Peschel_2003}
\begin{equation}
	\begin{split}
S(L,L/4)&=\sum_{i=1}^{L/4} s(\lambda_{i}),\\
s(\lambda_{i})&=-\lambda_{i} \text{log} \lambda_{i}-(1-\lambda_{i}) \text{log} (1-\lambda_{i}),
	\end{split}
	\end{equation}
where $\lambda_{i}$ are eigenvalues of sub-correlation matrix $G^{A}_{m,n}=\langle c^{\dagger}_{m}c_{n}\rangle$, $m, n\in A$.

Another quantity to characterize the strength of the skin effect is classical entropy \cite{MISE},
\begin{equation}
	\begin{split}
	S_{\text{cl}} =& -\sum_{i=1}^{L}\left[\langle n_{i}\rangle \text{log} \langle n_{i}\rangle+(1-\langle n_{i}\rangle) \text{log} (1-\langle n_{i}\rangle)\right] ,
		\end{split}
	\end{equation}
where $\langle n_{i}\rangle$ is the expectation value of particle number on site $i$ which can be obtained from the $i$-th diagonal elements 
of the correlation matrix 
$G_{i,i}=\langle\hat{c}^{\dagger}_{i}c_{i}\rangle$.
In addition, $S_\text{cl}$ is an upper bound of bipartite entanglement entropy. Due to the subadditivity of bipartite entanglement entropy \cite{nielsen_chuang, subadditivity}, the entanglement entropy of subsystem $A$ satisfies the following inequation
\begin{eqnarray}
S_{A}=\frac{1}{2}(S_{A}+S_{B})\leq \frac{1}{2}\sum_{i} S_{i},
\end{eqnarray}
where $B$ is the complementary subsystem with $A$, and $S_{i}$ is entanglement entropy of site $i$. In general, $\sum_{i} S_{i} \leq S_\text{cl}$. 
Therefore, the entanglement entropy $S(L, L/4)$ is bounded by half of $S_\text{cl}$. In this work, we focus on the half-filling case with the particle number $N=L/2$, and the evolution described by SSE respects $U(1)$ symmetry, i.e., the particle number is conserved.
With strong skin effect, about one-half of sites are occupied and one-half unoccupied (see Fig. \ref{suppfig3}), so apparently $S_\text{cl} \sim O(1)$ and thus $S_\text{cl}$ is independent of system size $L$. By contrast, in the absence of the skin effect, $S_\text{cl} \sim O(L)$.

\begin{figure}[htb]
\centering
\centering
\includegraphics[height=3.4cm,width=8.6cm]{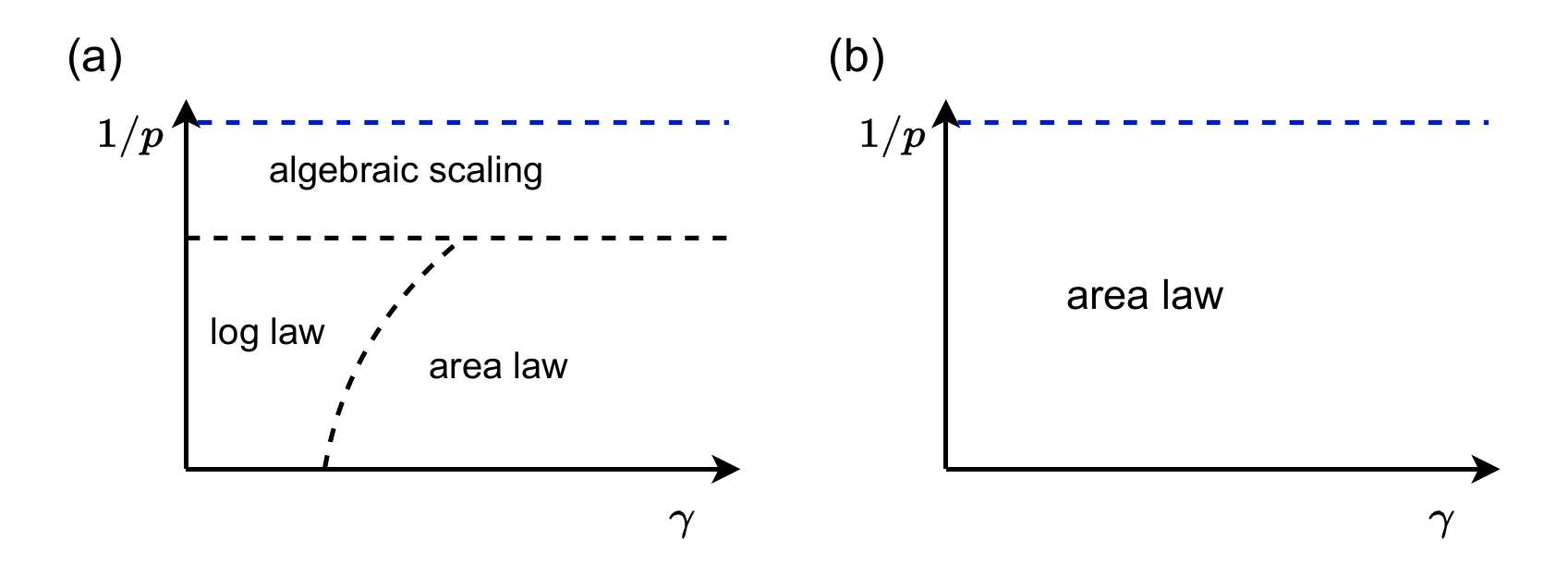}
\caption{Schematic phase diagram for $p>1$ (blue dotted line corresponds to $p=1$) and $\gamma>0$. (a) There are three phases: ``logarithmic law", ``area law," and ``algebraic scaling" phases (PBC). (b) The entanglement is suppressed by measurement-induced skin effect and obeys ``area law" (OBC).}
\label{phasediagram}
\end{figure}

\section{Numerical results and Phase diagram  }
\label{sec:longrange}

The model proposed above has two tunable parameters. One is monitoring rate $\gamma$, which controls the strength of unidirectional particle flows, i.e., the measurement-induced skin effect, caused by $\hat{H}_{\text{eff}}$ and quantum jumps. 
For OBCs, the measurement-induced skin effect is robust against the relatively short-range hopping and nearest interaction \cite{MISE}. 
Then Pauli exclusion principle freezes almost all particles, leaving fluctuations in a small region around the boundary of the occupied and 
unoccupied region (Fig. \ref{fig4}(b)). While for PBCs, particles flow around the bulk, and thus no skin effect appears (Fig. \ref{fig4}(a)). 
Via changing boundary conditions, we can explore the role of the skin effect.
The other tunable parameter is the hopping decay exponent $p$. With decreasing of $p$, the hopping range gets extended. In principle, long 
enough hopping will weaken or even eliminate the skin effect. Via tunning $\gamma$ and $p$ with OBC, we can see the competition between skin effect and long-range hopping.
It's worth noting that the generalized measurement is followed by conditional feedback, and we choose $\theta=\pi$ for simplicity
in the whole  Sec. \ref{sec:longrange}.

\subsection{Skin effect induced area law}

\begin{figure}[h]
\centering
\subfigure{
\includegraphics[height=3.5cm,width=4.0cm]{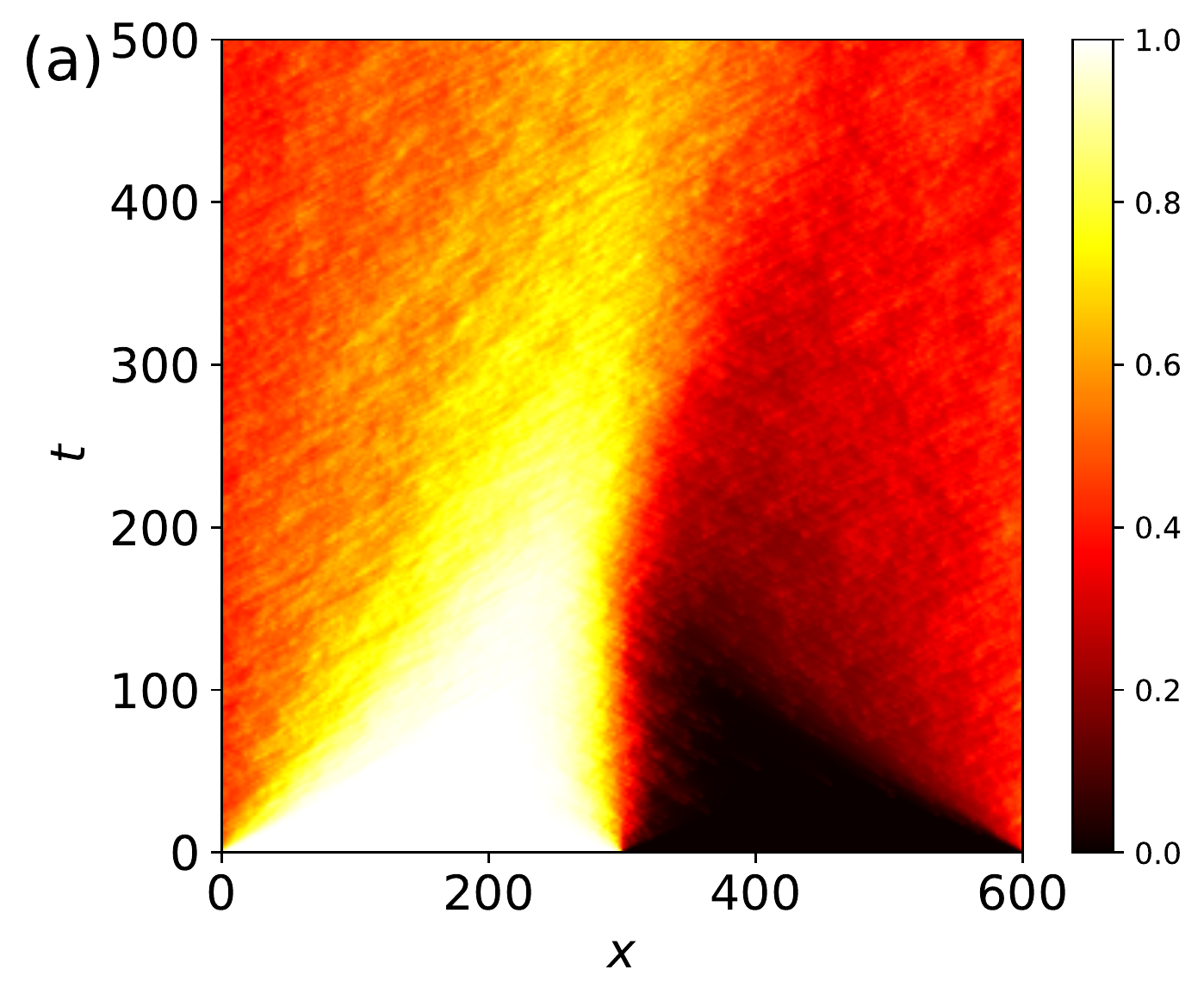}}
\subfigure{
\includegraphics[height=3.5cm,width=4.0cm]{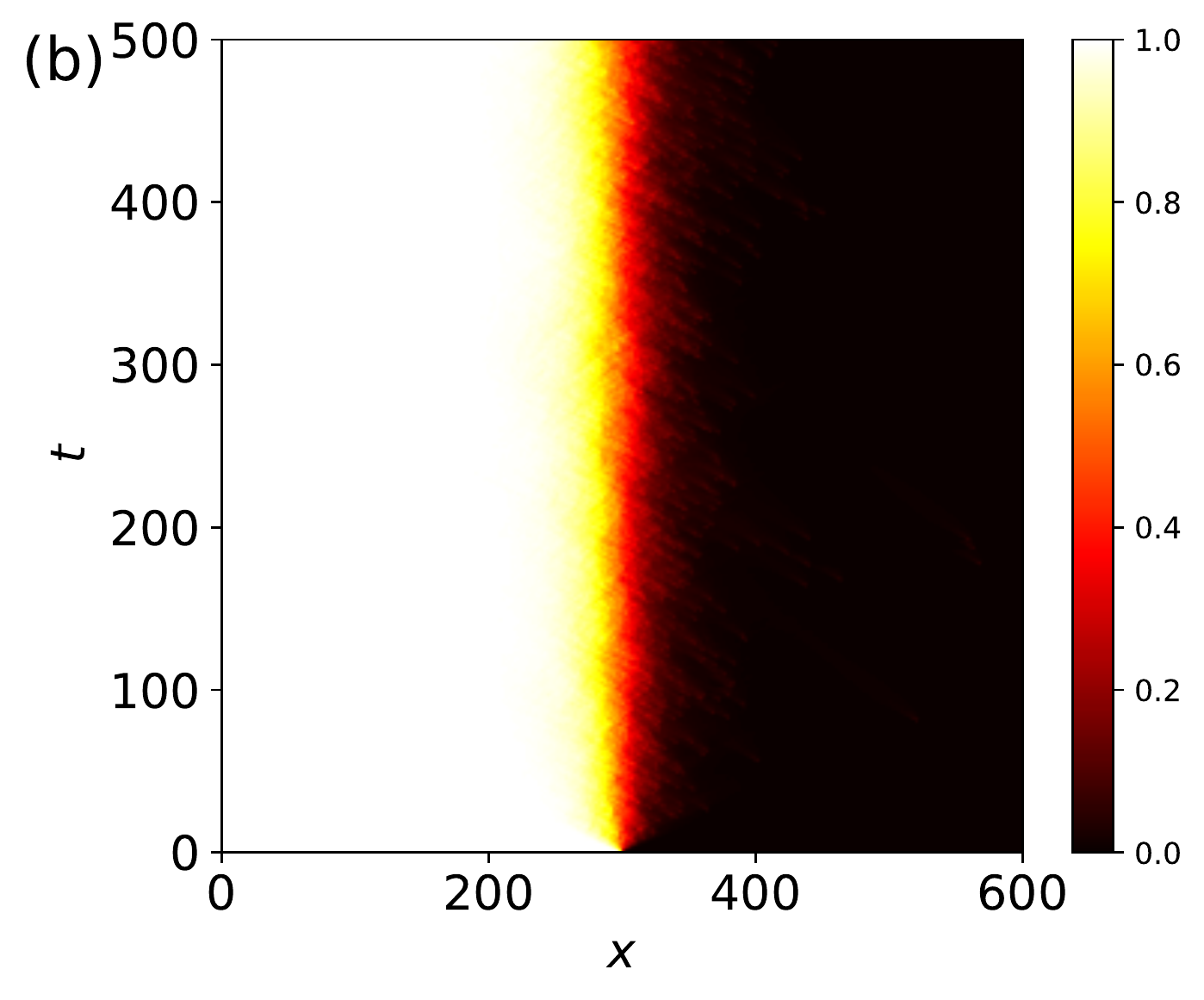}}
\caption{Density distribution evolution with time for $p=2.0, \gamma=0.1$. The initial state is chosen as $|111..000\rangle$. (a) PBC, no skin effect, particle number distribution of steady-state tends to be uniform. (b) OBC, with skin effect, steady state is almost half occupied and half unoccupied.}
\label{fig4}
\end{figure}

\begin{figure}[h]
\centering
\subfigure{
\includegraphics[height=3.5cm,width=4.0cm]{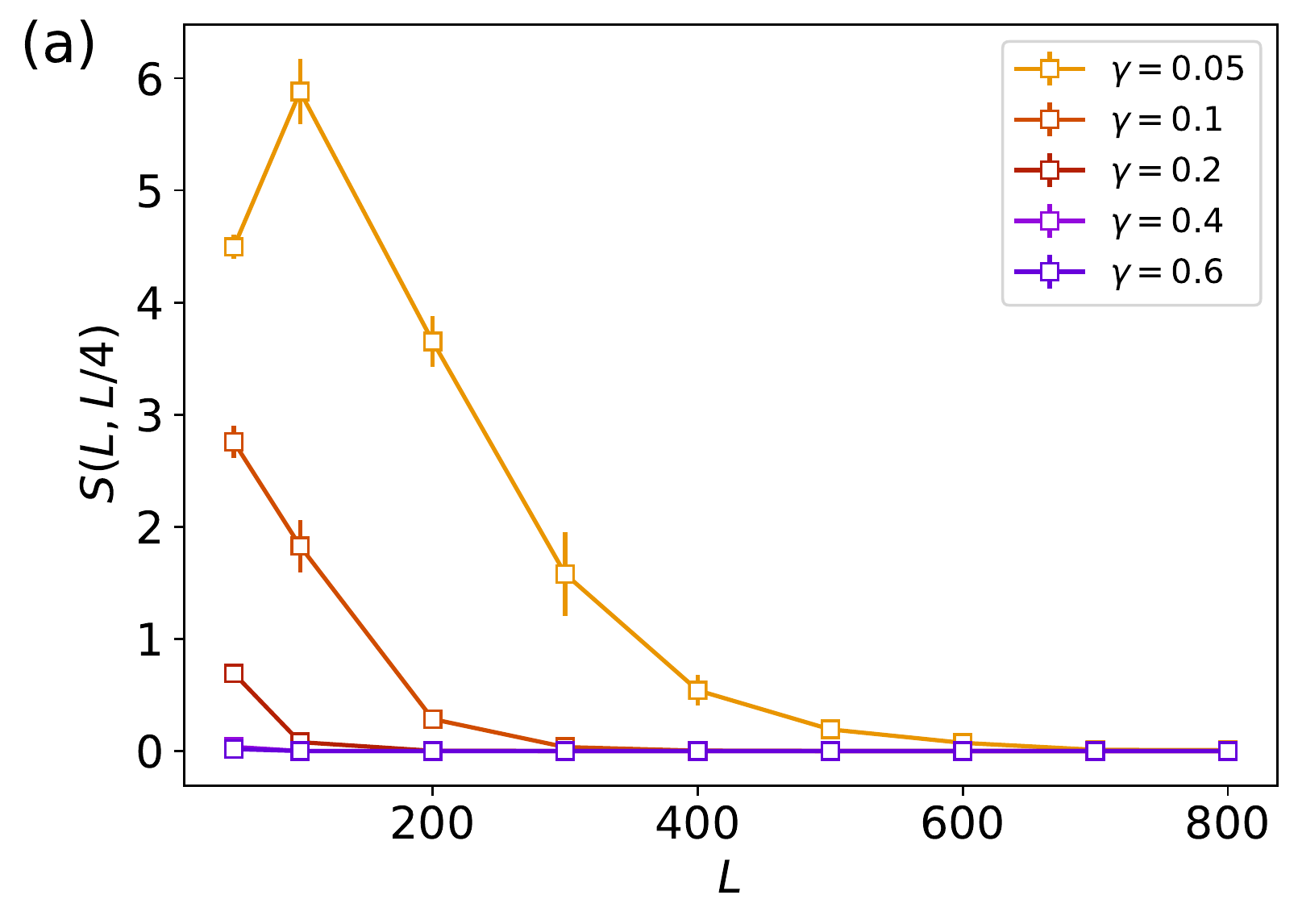}}\hspace{0.2em}%
\subfigure{
\includegraphics[height=3.5cm,width=4.0cm]{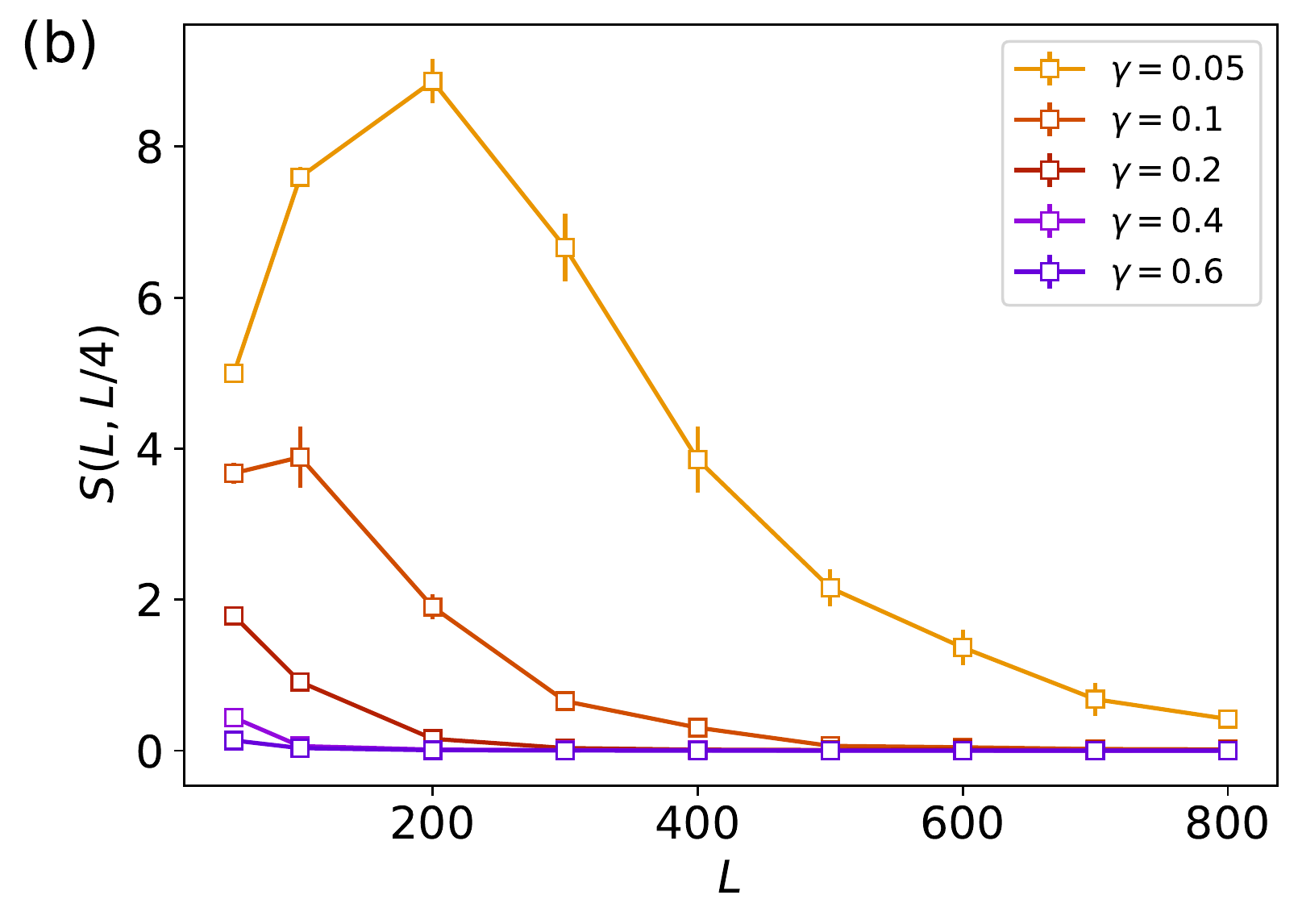}}
\subfigure{
\includegraphics[height=3.5cm,width=4.0cm]{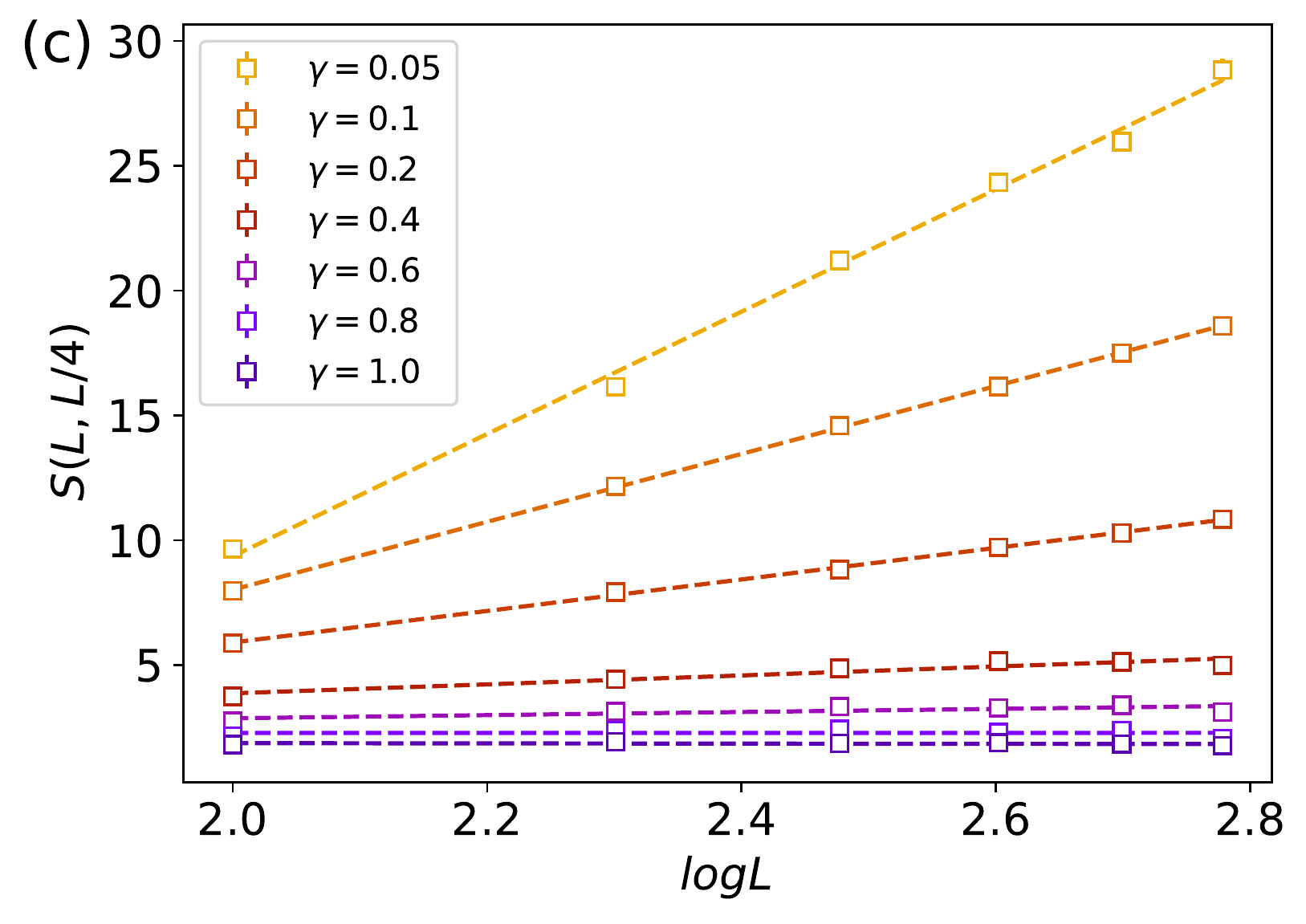}}
\subfigure{
\includegraphics[height=3.5cm,width=4.0cm]{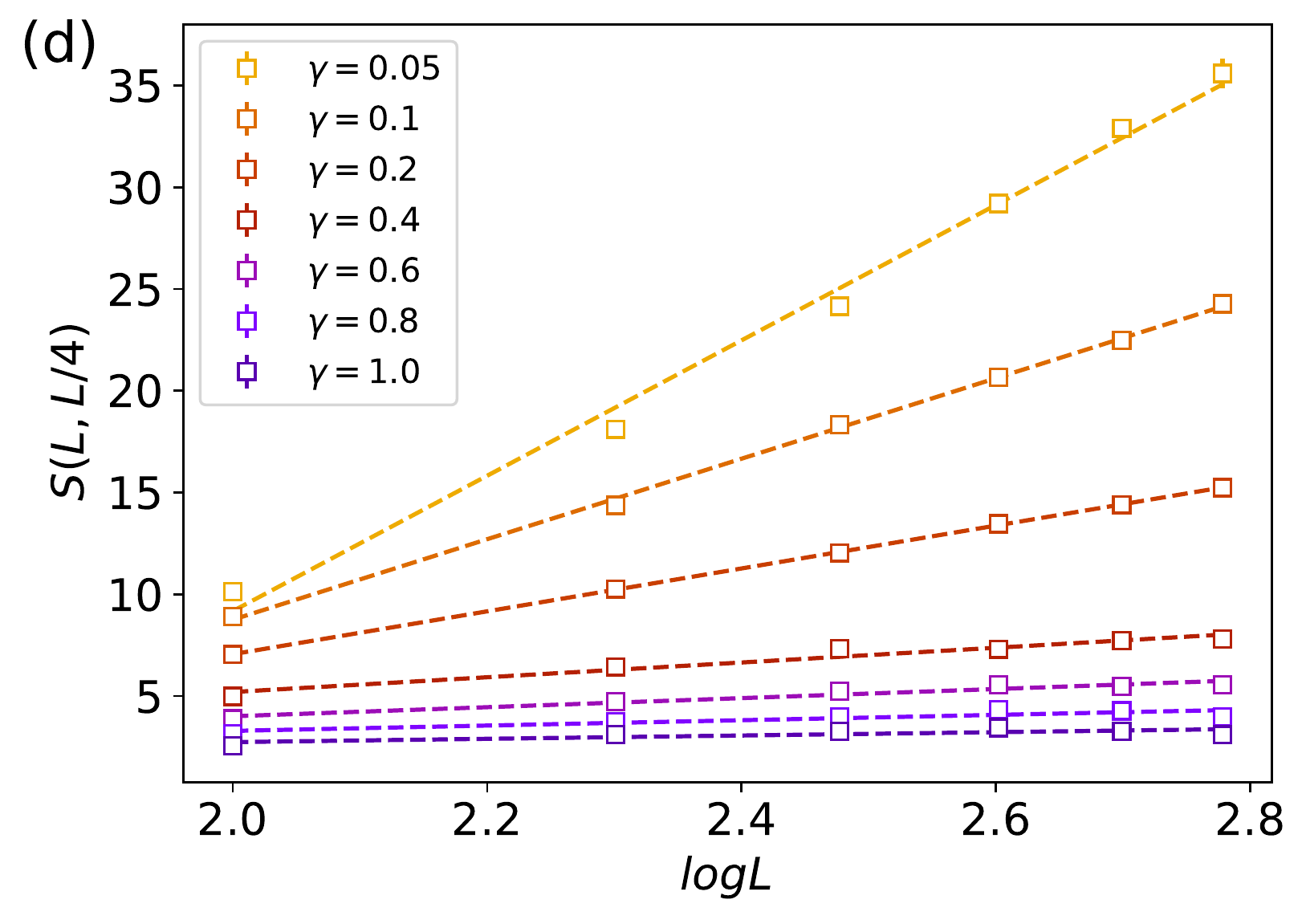}}


\caption{Entanglement entropy $S(L,L/4)$.  (a) OBC, $p=5.0$ (b) OBC, $p=2.0$. In both (a) and (b), $S(L, L/4)$ decays to zero, showing  the ``area law" behavior.  (c) PBC, $p=5.0$ (d) PBC, $p=2.0$.  In both (c) and (d), for small $\gamma$, $S(L, L/4)$ follows the ``logarithmic law"; for large $\gamma$, it follows the ``area law".}
\label{fig5}
\end{figure}
From the local density distribution, as shown in Fig. \ref{fig4}, the measurement-induced skin effect under OBC can be observed. When the fluctuation region is comparable with the system size, there exists quantum correlations between subsystems $A$ and $B$, and the bipartite entanglement entropy is finite. However, as the system size increases, subsystem $A$ is far away from the middle fluctuation zone, and the particles in $A$ are completely frozen, which greatly suppresses the entanglement. Therefore, when $L$ is larger than a threshold $L_{0}$, we can see that the entanglement entropy decay to zero fastly with increasing $L$ as shown in Fig. \ref{fig5}. This measurement-induced skin effect that suppresses the entanglement can be enhanced by more frequent generalized measurement (larger $\gamma$) and shorter range hopping (larger $p$) as shown in Fig. \ref{fig5}(a)(b) and 
Fig. \ref{Delta n}(a)(b). 

On the other hand, under PBCs, the skin effect disappears, and the local density distribution is uniform when the system reaches the steady state, as shown in Fig. \ref{fig4}(a). The entanglement entropy clearly shows the entanglement phase transition from the ``logarithmic law" phase into the ``area law" phase with the increase of measurement rate $\gamma$ as shown in Fig. \ref{fig5}(c)(d), which is consistent with the previous studies \cite{DiehlMIPT,longrangeMIPT}.

Based on the numerical results discussed above, we conclude that the skin effect will survive and induce the ``area law'' entanglement phase with relatively long-range hopping ($p \ge 2.0$) and monitoring rate  $\gamma \geq 0.05$. In the next subsection, we will draw the same conclusions 
from the scaling relation of the classical entropy $S_\text{cl}$ and further get some intuition for the case $1 < p < 2.0$ and 
$\gamma \rightarrow 0$.


\subsection{Scaling behavior of $S_\text{cl}$}

\begin{figure}[h]
\centering
\centering
\subfigure{
\includegraphics[height=3.5cm,width=8.4cm]{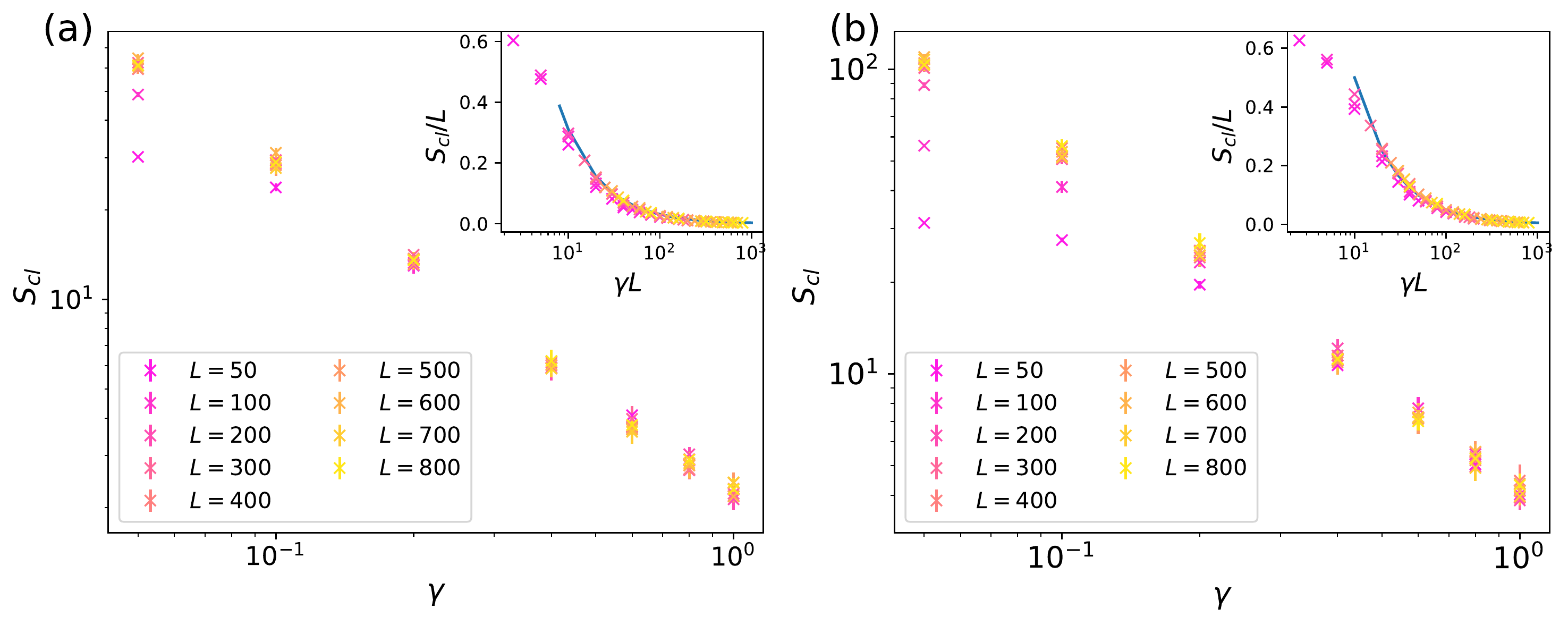}}
\caption{Classical entropy $S_{\text{cl}}$ versus $\gamma$ on the log-log scale for different system sizes. (a) $p=5.0$, the blue curve in the inset shows the fit $y=3.1/x$ for large $\gamma L$. (b) $p=2.0$, the blue curve in the inset is the fit $y=5/x$ for large $\gamma L$.
The scaling relation tells that, even for small $\gamma$, $S_\text{cl}$ behaves as $S_\text{cl}\propto 1/\gamma$ in the thermodynamics limit.}
\label{fig6} 
\end{figure}

To further demonstrate the area law phase at OBCs for any $\gamma >0$, we study the finite-size scaling behavior of the classical entropy 
$S_\text{cl}(L)$.  

As illustrated in the insets of Fig. \ref{fig6} (a) and (b) for large $p\ge 2.0$, we find the  finite-size scaling relation proposed in \cite{MISE}
\begin{equation}
	S_\text{cl}(L) =L f(\gamma L),
	\label{fss}
\end{equation}
still holds well for a large range of $p$ we calculated.

In the limit $\gamma \to 0$ and finite $L$, this relation holds with $f(0)$ a constant, considering the extension of $S_\text{cl}$.
Physically, for $\gamma$ bigger than a threshold $\gamma_{0}$, the skin effect is so strong that the size of fluctuation areas is 
stabilized (about $O(1)$) for medium system size ($L\sim 10^2$). Therefore, one expects that $S_\text{cl}(L)$ be invariant with system size $L$,
as $L \to \infty$.
The scaling Eq. (\ref{fss}) then requires the scaling function 
$f(\gamma L) \propto 1/(\gamma L)$ as $\gamma L \rightarrow \infty $, which is consistent with the fits for large $\gamma L$ shown in 
the insets of Fig. \ref{fig6}. Hence, 
in the thermodynamics limit, for $\gamma > 0$, $S_\text{cl}(\gamma,L\rightarrow\infty)\propto 1/\gamma$.  Therefore, 
for any nonzero $\gamma$, the entanglement entropy is bounded in the thermodynamics limit,
and the system enters the ``area law'' entanglement phase immediately in the presence of generalized measurement ($\gamma>0$).

\begin{figure}[h]
\centering

\includegraphics[height=6.5cm,width=8.2cm]{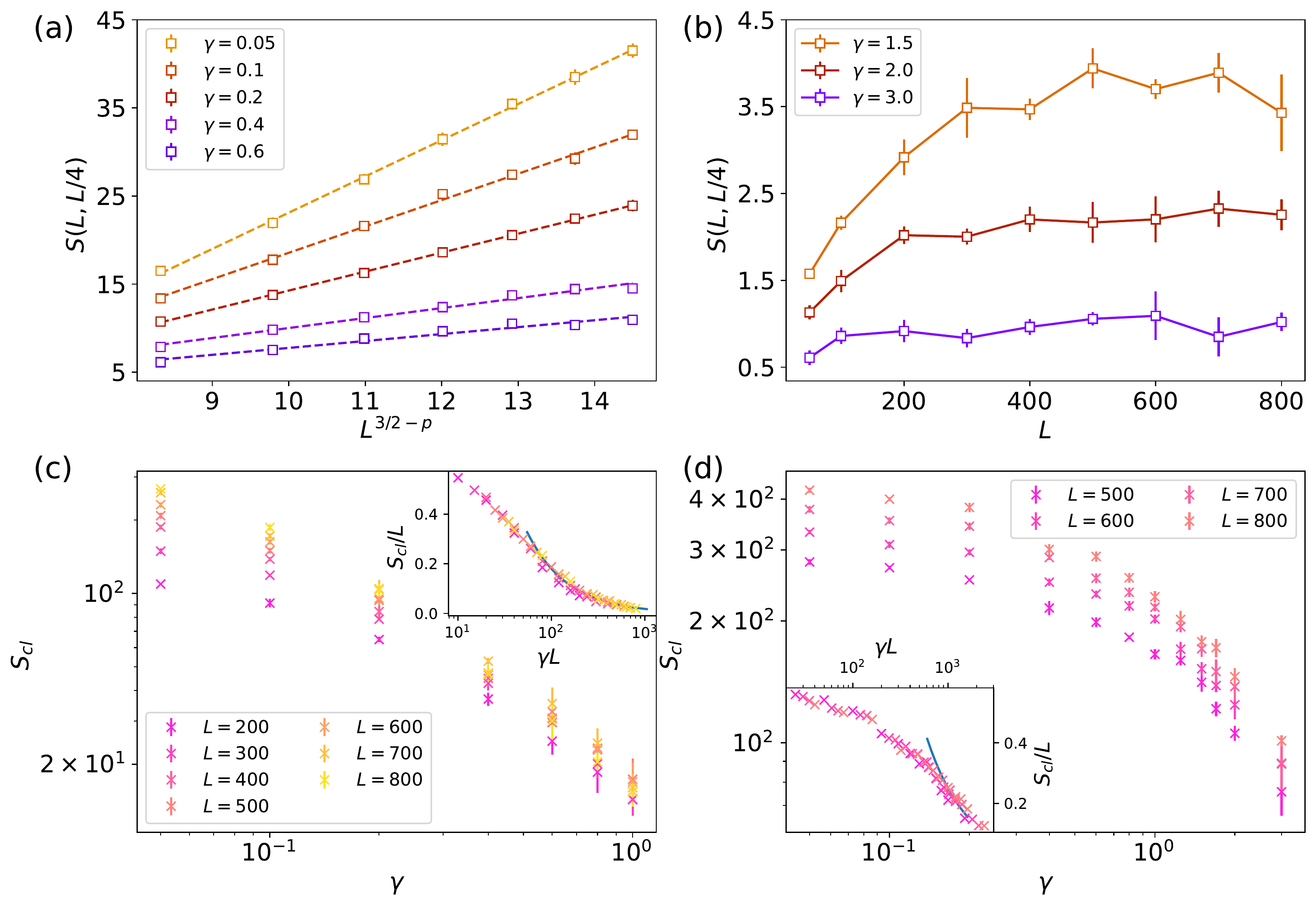}
\caption{Entanglement entropy and classical entropy. (a) $p=1.1$, for small $\gamma$ and medium size ($L\sim 10^2$), entanglement entropy $S(L, L/4)$ fits well with $L^{3/2-p}$. (b) $p=1.1$, for large $\gamma$, $S(L, L/4)$ quickly saturates. (c) Classical entropy $S_\text{cl}$ and its data collapse for $p=1.5$. The blue fitting curve for large $\gamma L$ in the inset is $y=18/x$. (d) Classical entropy $S_\text{cl}$ and its data collapse for $p=1.1$. The blue fitting curve for large $\gamma L$ in the inset is $y=250/x$.}
\label{fig8}
\end{figure}

Worth to mention numerical results show that the threshold $\gamma_{0}$ increases as $p$ lowers down (see Fig. \ref{fig6} and Fig. \ref{fig8}). 
When $p$ is close to 1, the threshold $\gamma_{0}$ gets so big and beyond our numerical capabilities. For example, as Fig. \ref{fig8}(c), (d) 
show, in the range of $\gamma$ we calculated, when $p=1.5$ and $p=1.1$, $S_\text{cl}(L)$ does not collapse well for system sizes $L\leq 800$. 
However, there is a clear tendency that for bigger $\gamma$, although beyond our numerical abilities,  $S_\text{cl}(L)$ for various system sizes $L$  will ultimately collapse.  Moreover, according to the data collapse in the inset of Fig. \ref{fig8} (c), (d), we see 
Eq. (\ref{fss}) is still satisfied, and for $\gamma L\rightarrow \infty$, $f(\gamma L)$ tends to behave as $1/\gamma L$ like in Fig. \ref{fig6}, which 
means in the thermodynamical limit, $S_\text{cl} \propto 1/\gamma$. 

The behaviors of $S_\text{cl}$ for different $p$ are consistent with $\Delta n$. As shown in Fig. \ref{Delta n}, when $p=2.0, 5.0$,  for moderate $\gamma$ and medium size $L$, $\Delta n$ is close to 1, which means particles almost completely localize at the left side, thus $S_\text{cl}$ should be independent of $L$ and data of $S_\text{cl}$ for various sizes collapse well. While for $p=1.1$, even for $L=800$ and $\gamma=3.0$, $\Delta n$ is still not close to 1, which means $S_\text{cl}$ will be dependent on $L$ and data of $S_\text{cl}$ for various sizes collapse badly. However, according to Fig. \ref{Delta n}, it's also reasonable to predict for larger $L$ and bigger $\gamma$, $\Delta n$ will finally close to 1, thus $S_\text{cl}\sim O(1)$. Therefore, for $\gamma L\rightarrow\infty$, $S_\text{cl} \propto 1/\gamma$ always holds.




The increase of threshold $\gamma_{0}$ is also observed for the nearest hopping case  ($p=\infty$) \cite{MISE}. By changing the $\theta$ of the unitary feedback phase factor, it was found that when $\theta$ decreases, the threshold $\gamma_{0}$ increases \cite{MISE}. This is understandable since the decrease of $\theta$ and $p$ weakens the skin effect and extends the fluctuation areas. Therefore, one needs bigger $\gamma$, namely more 
frequent measurements to strengthen the skin effect and reduce fluctuation areas.

\subsection{Finite-size algebraic scaling }

\begin{figure}[h]
\centering
\subfigure{
\includegraphics[height=3.0cm,width=4.0cm]{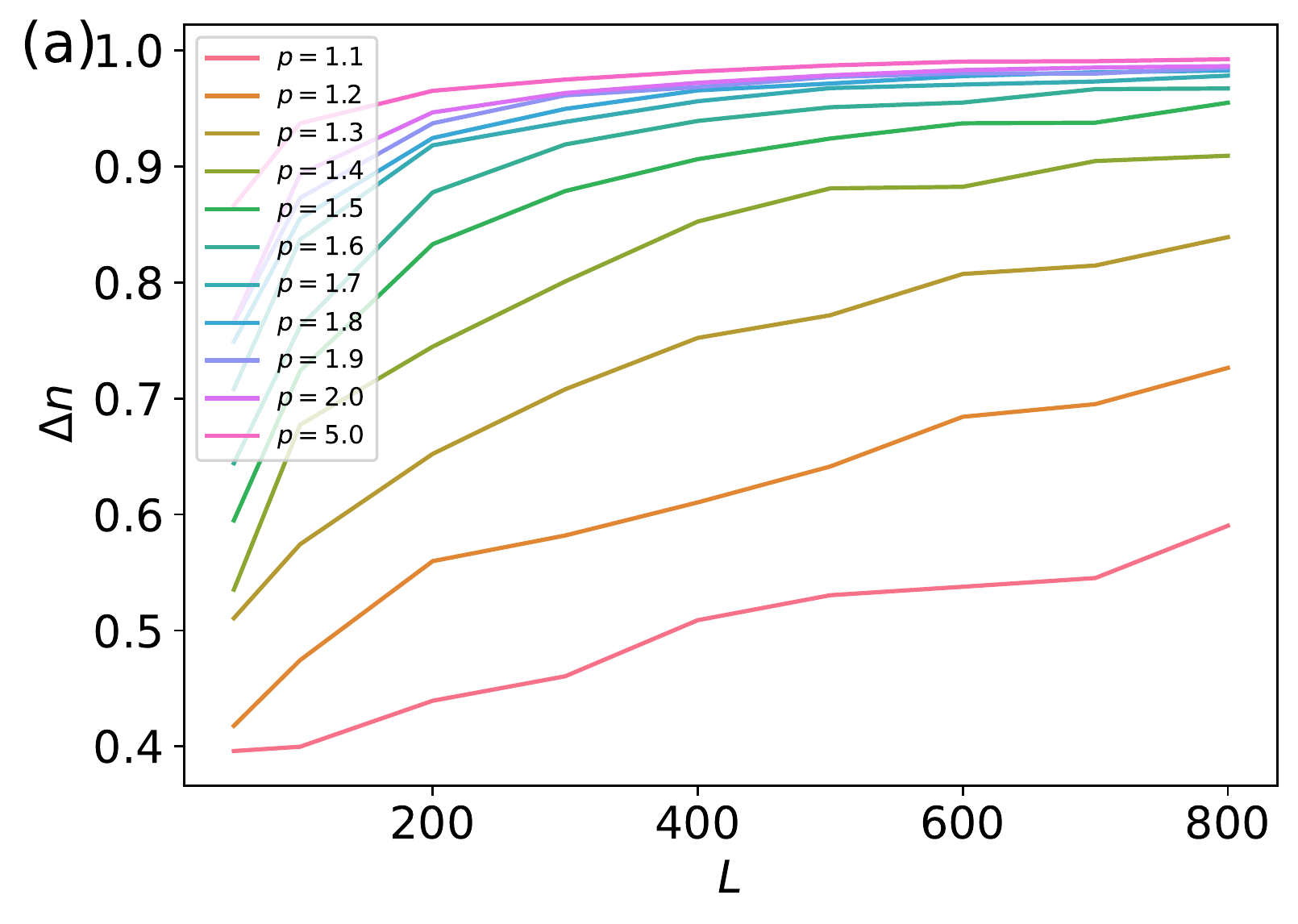}}
\subfigure{
\includegraphics[height=3.0cm,width=4.0cm]{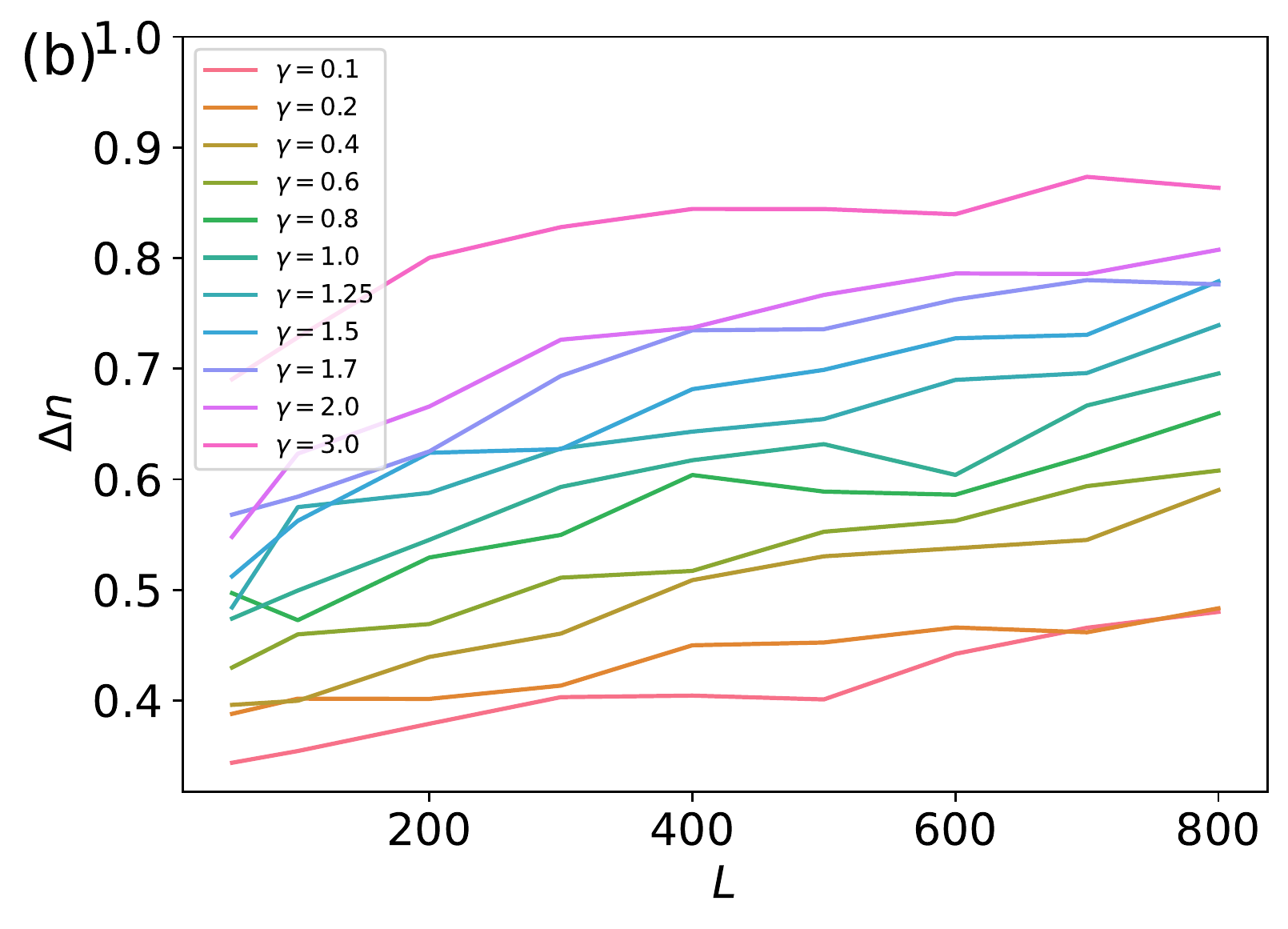}}
\caption{The imbalance of the particle number distributions between the left and right half side. (a) $\gamma=0.4$, with the decrease of hopping decay exponent $p$, $\Delta n$ decreases, skin effect gets weaker. While with the increase of system size $L$, $\Delta n$ grows, namely skin effect gets stronger. (b) $p=1.1$, with the increase of monitoring rate $\gamma$ and system size $L$, $\Delta n$ grows and skin effect gets strengthened. }
\label{Delta n}
\end{figure}

It's known that, for $p\ge 2.0$ and nonzero $\gamma$, the skin effect already dominates for medium system size ($L \sim 10^2$). Further 
decreasing $p$, we find that long-range hopping effects emerge at least for medium system size ($L\sim 10^{2}$) when both $p$ and $\gamma$ 
are relatively small. Take $p=1.1$ as an example, as shown in Fig. \ref{fig8}(a), for small $\gamma$ and system sizes $L$ up to 800, although the 
skin effect reduces the entanglement overall, the entanglement entropy $S(L, L/4)$ seems to scale as $L^{3/2-p}$, similar to what found in \cite{longrangeMIPT}. However, this algebraic scaling behavior is a finite-size effect.

In the following, we will support our viewpoint from two aspects. The first aspect is about particle number distributions. For medium size 
	systems ($L\sim 10^{2}$), the entanglement entropy is mainly contributed from the middle fluctuation area, which is comparable with system 
 size $L$ due to long-range hopping. However, with the increase of $L$, the size of the fluctuation zone narrows down with respect to system size 
 $L$ (see Fig. \ref{suppfig2}). This is consistent with Fig. \ref{Delta n}(b). With the increase of $L$, the proportion of particles in the left half side grows, and the skin effect strengthens. So in the thermodynamics limit, it's reasonable to argue that the fluctuation areas are in the order of $O(1)$ size; thus, the entanglement entropy will saturate. The second aspect is from the scaling behavior Eq. (\ref{fss}) of $S_\text{cl}(L)$. As 
 discussed in the previous subsection, even for $p\rightarrow 1$, the classical entropy $S_\text{cl}(\gamma)$ still behaves as $1/\gamma$ for any 
 nonzero $\gamma$ in the thermodynamic limit, so the entanglement entropy should be bounded. Therefore, based on these two aspects, the algebraic 
 scaling behavior should disappear in the thermodynamic limit. 

\begin{figure}[h]
\centering
\subfigure{
\includegraphics[height=3.5cm,width=4.0cm]{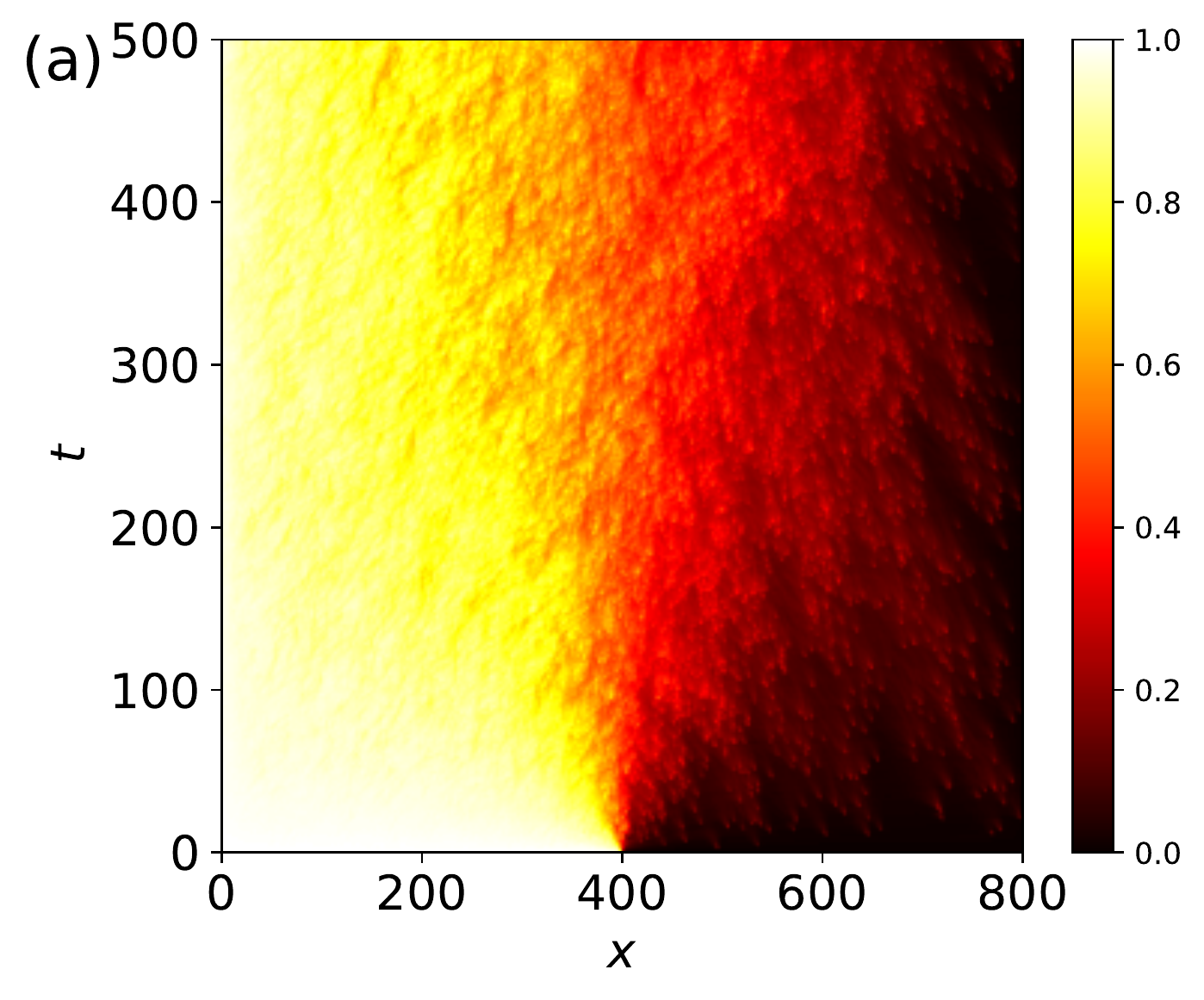}}
\subfigure{
\includegraphics[height=3.5cm,width=4.0cm]{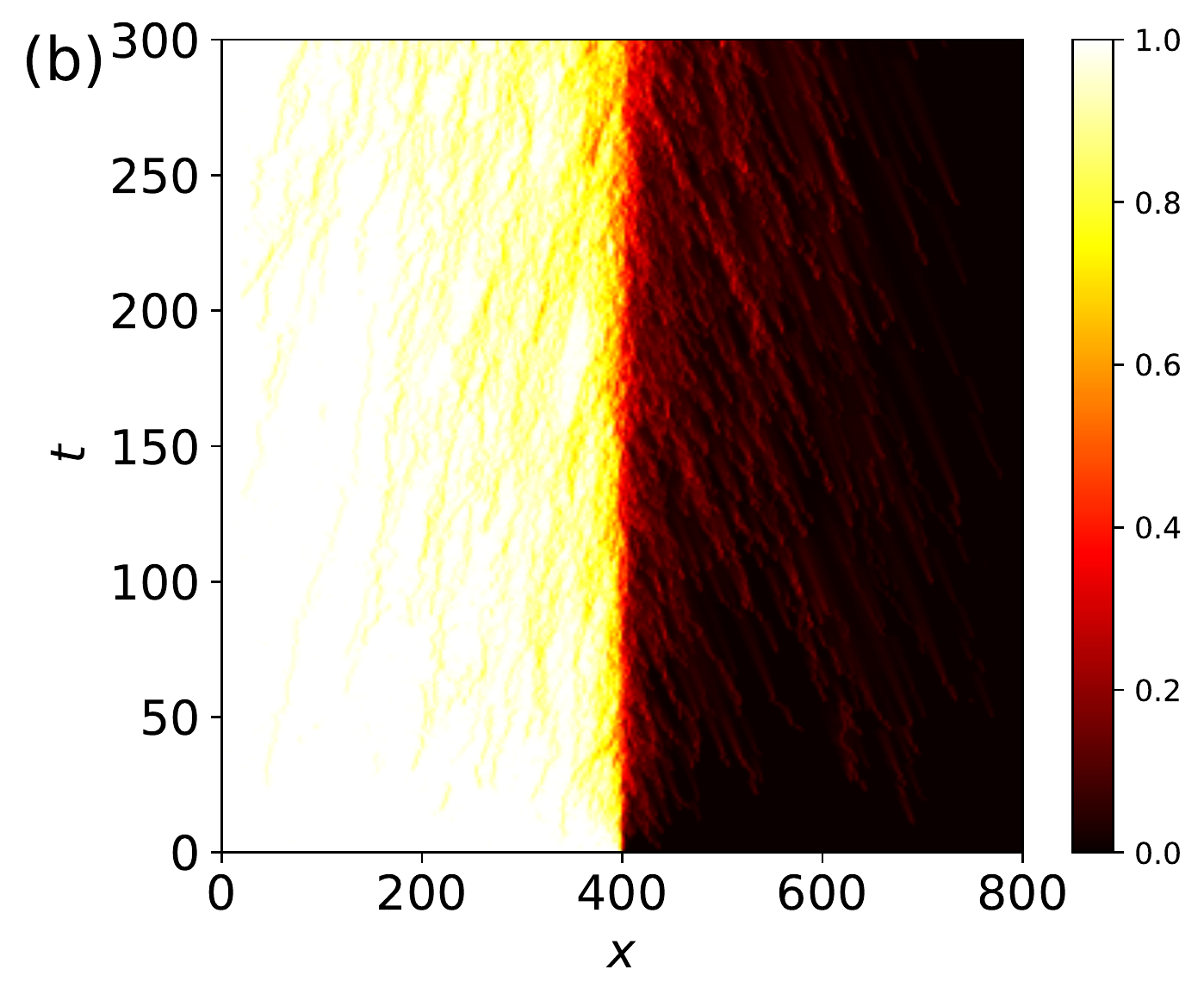}}
\caption{Density distribution evolution with time for $p=1.1$ and $L=800$ under OBCs, the initial state is chosen as $|111..000\rangle$. (a) $\gamma=0.1$, skin effect is weak. Quantum correlation is developed in the middle zone ($\sim O(L)$ size). (b) $\gamma=3.0$, skin effect is strong. }
\label{fig7}
\end{figure}

	Indeed, when the monitoring rate $\gamma$ is big, as shown in Fig. \ref{fig8}(b), the entanglement entropy quickly saturates with $L$. These 
	entanglement entropy behaviors are also in agreement with particle number distributions. As shown in Fig. \ref{fig7}(a) and \ref{fig7}(b), 
	at $p=1.1$, for small $\gamma$, the skin effect is weak, the quantum correlation zone is comparable with system size; thus finite-size algebraic scaling behaviors emerge. While for large $\gamma$, unidirectional flow increases, thus the skin effect gets strengthened, which greatly suppresses entanglement growth.

	So the ``logarithmic law" and ``algebraic law"  phases are absent due to the skin effect and leaving only the ``area law" phase. In other words, power-law hopping with exponent $p>1$ improves the size of fluctuation areas and weakens the skin effect but can not completely eliminate the skin effect. Therefore, in the thermodynamic limit, the skin effect will always dominate, thus induce ``area law" phase. 

The schematic phase diagrams for PBCs and OBCs are depicted in Fig. \ref{phasediagram}. For the PBCs, the phase diagram is consistent with previous 
results \cite{longrangeMIPT}, in which the ``observer'' continuously measures the local particle density. Theoretically, for $p\leq3/2$ ($p\geq3/2$), 
the long-range hopping is relevant (irrelevant), which is reflected on the first-order RG equation \cite{longrangeMIPT}. Therefore, for 
$p\leq 3/2$, frequent local measurements cannot overcome the entanglement generated by long-range hopping. This theoretical analysis also fits 
the PBCs case here since the measurements in our model only include local density measurement and nearest hopping; thus, the long-range hopping 
shall still dominate for $p\leq 3/2$. However, the analysis breaks down for the OBCs due to the skin effect. Because most particles tend to 
localize at one side and be next to each other under OBCs, the Pauli exclusion principle hinders long-range hopping. Therefore, even for 
$p\leq 3/2$, the long-range hopping is suppressed, leading to the ``area law" phase.

\section{No-feedback case}
\label{sec:nofeedback}

\begin{figure}[h]
\centering
\includegraphics[height=6.6cm,width=8.6cm]{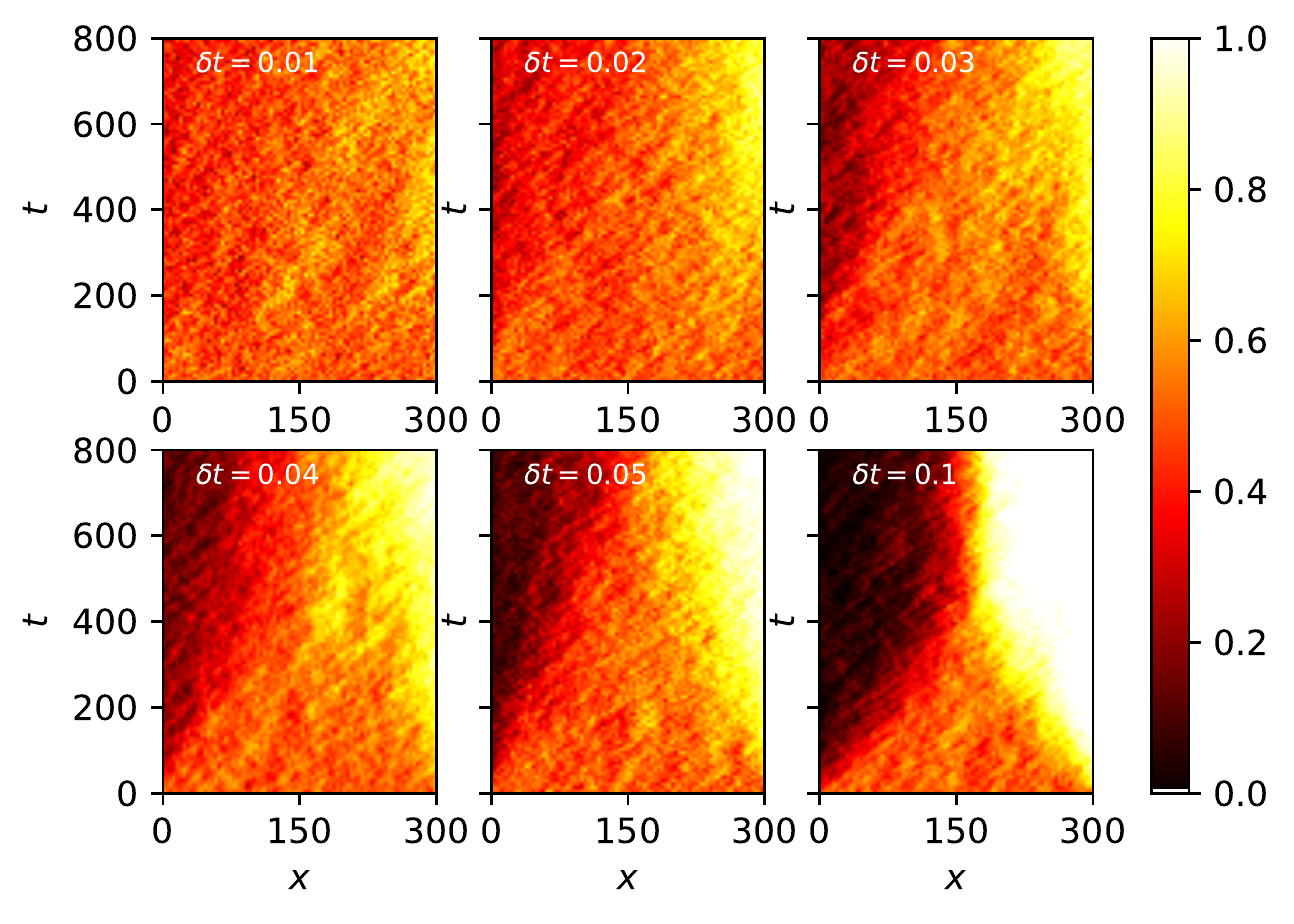}
\caption{Density distribution evolution with time under OBCs for the no-feedback case with $\gamma=2.0$, $p=2.0$, $L=300$. The initial state is $|1010...10\rangle$. The six subfigures correspond to time step $\delta t=0.01, 0.02, 0.03, 0.04, 0.05, 0.1$. When $\delta t$ is small enough, there is no skin effect; while increasing $\delta t$, the ``pseudo skin effect" emerges.}
\label{densitynofeedback}
\end{figure}

The previous section (Sec. \ref{sec:longrange}) focuses on the case with conditional feedback ($\theta=\pi$). Here, we will investigate no 
conditional feedback case ($\theta=0$) briefly. It's known that the density matrix, which can be acquired through averaging over all measurement outcomes (trajectories), evolves according to the Lindblad master equation \cite{lindblad, OQS}
\begin{equation}
	\begin{split}
	\dfrac{d\hat{\rho}}{dt} =& -i\left[\hat{H}, \hat{\rho}\right]+\gamma\sum_{\alpha}\big(\hat{L}_{\alpha}\hat{\rho}\hat{L}_{\alpha}^{\dagger}-\dfrac{1}{2}\{\hat{L}_{\alpha}^{\dagger}\hat{L}_{\alpha},\hat{\rho}\}\big) .
		\end{split}
	\end{equation}
 For no-feedback case, the Lindblad operator $\hat{L}_{i}=\dfrac{1}{2}\hat{\xi}_{i}^{\dagger}\hat{\xi}_{i}$ is Hermitian.
 Apparently, the steady state is the maximally mixed state $\rho_{ss} \sim\Bbb I$, thus the particle distribution is uniform since the number 
 operator $\hat{n}_{i}$ is linear. Therefore, there is no skin effect for the no-feedback case, and  theoretically, we can observe MIPT for the no-feedback case under OBCs. However, numerically we find the quantum jump simulation method is sensitive to time step $\delta t$ for the no-feedback case when $\gamma$ is relatively large. As Fig. \ref{densitynofeedback} shows, with the increase of $\delta t$, the ``pseudo skin effect" 
 emerges and destroys MIPT. 
 
\begin{figure}[h]
\centering

\includegraphics[height=5.2cm,width=7.5cm]{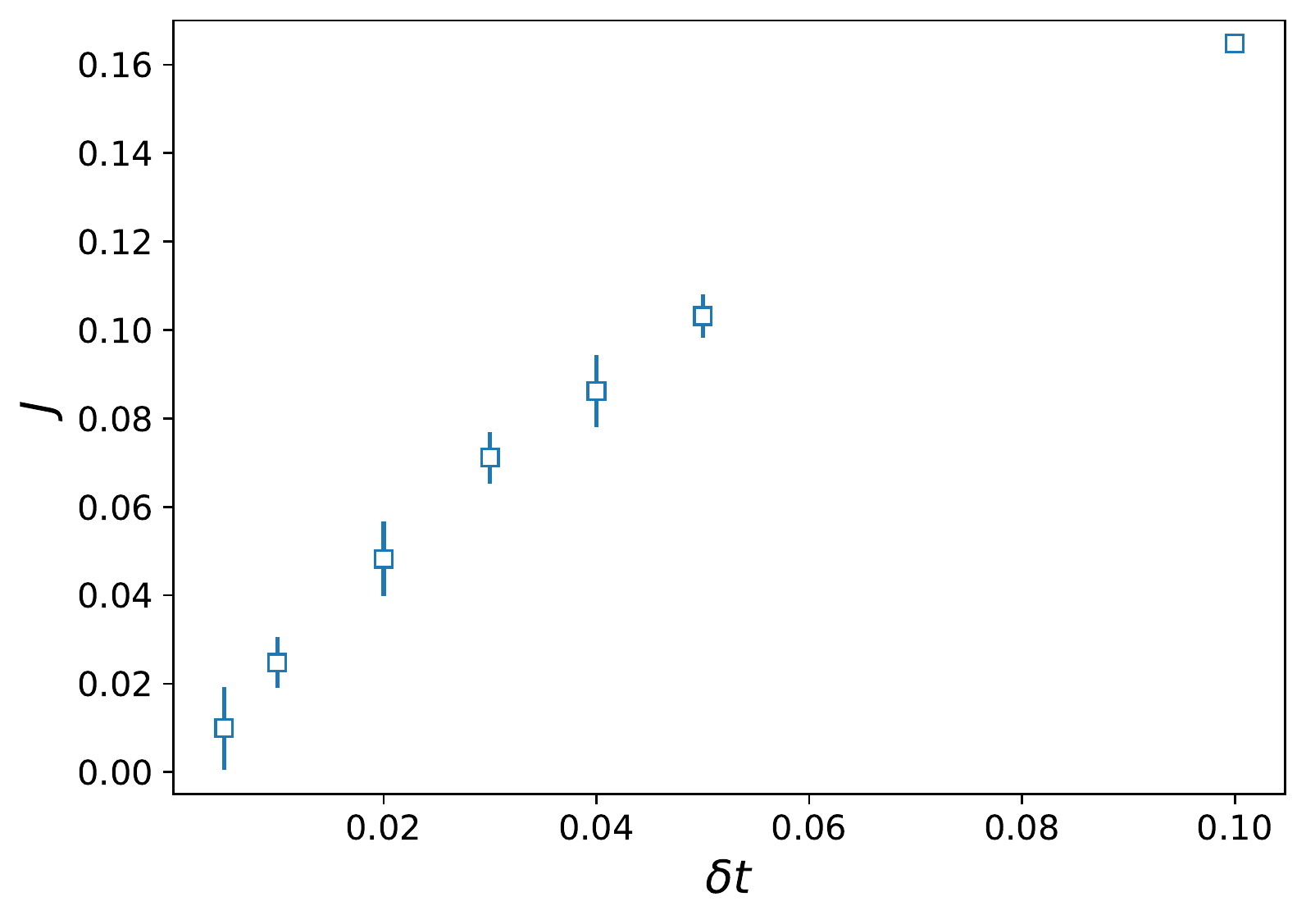}
\caption{The current $J$ under PBCs for the no-feedback case with $p=2.0$, $\gamma=2.0$, $L=300$. With the decrease of simulation time step 
$\delta t$, current $J$ decays to zero, which means that there is no skin effect in the continuous measurement limit $\delta t \rightarrow 0$.}
\label{currentnofeedback}
\end{figure}

 This phenomenon is also verified in particle current $J$ under PBCs, with $J$ defined as 
\begin{equation}
	\begin{split}
	J =& \dfrac{i}{L}\sum_{n=1}^{L}(\langle\hat{c}_{n}^{\dagger}\hat{c}_{n+1}\rangle-\langle\hat{c}_{n+1}^{\dagger}\hat{c}_{n}\rangle).
		\end{split}
	\end{equation}
Similar to the previous study of NHSE and Liouvillian skin effect \cite{WindingSkin, Liouvillianskin}, the nonzero PBC current means that the skin effect exists under OBCs. As demonstrated in Fig. \ref{currentnofeedback}, for $p=2.0$, $\gamma=2.0$, $L=300$, when $\delta t$ is not small enough, such as $\delta t =0.05$, the current $J$ is finite, and skin effect seems to emerge under OBCs. However, with the decrease of $\delta t$, the current $J$ vanishes. Therefore, there must be no skin effect in the continuous measurement limit for the no-feedback case, which is in agreement with the intuition from the Lindblad master equation. Theoretically, the errors caused by finite $\delta t$ are the consequences of utilizing the first-order approximation of $\gamma \delta t$ to effectively simulate continuous measurements. Therefore, we must carefully select sufficiently small $\delta t$ 
for big monitoring rate $\gamma$
to avoid the ``pseudo skin effect".

\section{Conclusion and Outlook}
\label{sec:conclsn}

In this paper, we proposed a monitored power-law long-range hopping free fermion model with generalized measurement and conditional feedback. Based on 
convincing numerical results, we have demonstrated that the measurement-induced skin effect (with OBCs) suppresses the entanglement within the 
system and thus drives the system to enter the ``area law'' entanglement phase. We have found that when the power-law decay exponent $p$ is 
large, i.e., relatively short-range hopping, the entanglement decays to zero fastly with increasing system size in the presence of the 
measurement, similar to the previous study with the nearest 
neighbor hopping \cite{MISE}. On the other hand, when the power-law decay exponent $p$ is small, we have found that the effect of long-range 
hopping dominates and observed an algebraic scaling for entanglement with the system size accessible in this work. However, as indicated by the 
local density distribution 
and classical entropy, we have shown that the systems ultimately enter the skin effect induced ``area law'' entanglement phase in the 
thermodynamic limit. 
Moreover, we have demonstrated that conditional feedback is still necessary for the measurement-induced skin effect with long-range hopping, 
Although the numerical results are sensitive to the time step $\delta t$. We further proved the existence of a ``pseudo skin effect"  for the case 
without conditional feedback.

In principle, the all-to-all hopping, i.e., the $p \rightarrow 0$ limit, can suppress the skin effect and may sustain large-scale entanglement. One 
interesting direction for future work is to explore other long-range hopping models which may exhibit the measurement-induced entanglement phase transition. In addition, the many-body interaction is unavoidably for natural physical systems. Another interesting direction is to explore the competition between skin effect and long-range interaction.

\begin{acknowledgments}

We thank Yupeng Wang and Jie Ren for the valuable discussions. This work was supported by the National Natural Science Foundation of China under Grant No.~12175015 and No.~11734002. 
The authors acknowledge the support extended by the Super Computing Center of Beijing Normal University.
\end{acknowledgments}

\bibliographystyle{apsreve}
\bibliography{ref}

\clearpage
\appendix
\renewcommand{\theequation}{S\arabic{equation}}
\setcounter{equation}{0}
\renewcommand{\thefigure}{S\arabic{figure}}
\setcounter{figure}{0}
\section{Details to simulate single trajectory and averaged trajectories dynamics}
\label{sec:appendixA}

Since the evolution equation (Eq. \ref{eq2}) is quadratic, if the initial state is Gaussian, it will preserve gaussianity through evolution. Assume the system size is $L$, particle number is $N$, for free fermions, state $|\psi\rangle$ can be written as $|\psi\rangle=\coprod\limits_{i=1}^N(\sum\limits^{L}_{j=1}U_{jl}(t)c^{\dagger}_{j})|0\rangle$.  $|\psi\rangle$ is a Slater determinant state of $N$ fermions, where the columns of $U$ give the single-particle wave functions. The state $|\psi\rangle$ can be simply represented by $L \times N$ matrix $|U\rangle$. 
Physically, the state will keep invariant with elementary column operations.

To simulate the stochastic Schrodinger equation (Eq. \ref{eq2}), it's easy to prove in the first order approximation of $\delta t$, the Eq. \ref{eq2} is equivalent to
\begin{equation}\label{S1}
	\begin{split}
	|\psi(t+\delta t)\rangle =dW_{n} \hat{L}_{n}\dfrac {e^{-i\hat{H}_\text{eff}\delta t}|\psi(t)\rangle}{\parallel \hat{L}_{n}e^{-i\hat{H}_\text{eff}\delta t}|\psi(t)\rangle\parallel}\\
    +(1-dW_{n})\dfrac {e^{-i\hat{H}_\text{eff}\delta t}|\psi(t)\rangle}{\parallel e^{-i\hat{H}_\text{eff}\delta t}|\psi(t)\rangle\parallel} ,
		\end{split}
	\end{equation}
in which $dW_{n}=\gamma\langle L^{\dagger}_{n}L_{n}\rangle\delta t$.	
Physically, the Eq. \ref{S1} is in accordance with quantum trajectory theory \cite{quantumtrajectory}, the state will evolve  under effective non-Hermitian Hamiltonian in a short time, then undergo possible quantum jumps and so on.

1) non-Hermitian evolution
\begin{equation}
	\begin{split}
	|\psi(t+\delta t)\rangle &= e^{-i\hat{H}_\text{eff}\delta t}|\psi(t)\rangle\\
	&= \prod\limits_{i=1}^N(\sum^L_{j=1}U_{jl}(t)e^{-i\hat{H}_\text{eff}\delta t} c^{\dagger}_j e^{i\hat{H}_\text{eff}\delta t})|0\rangle,
		\end{split}
	\end{equation}

Assume $\hat{H}_\text{eff}=\sum_{m,n}h^{mn}_\text{eff}\hat{c}^{\dagger}_{m}\hat{c}_{n}$,
using Baker–Campbell–Hausdorff formula, we can get, 
$e^{-i\hat{H}_\text{eff}\delta t} c^{\dagger}_j e^{i\hat{H}_\text{eff}\delta t} = \sum_{m=1}^{L}\left[ e^{-ih_\text{eff}\delta t}\right]_{m,j}c^{\dagger}_m$, 
\begin{equation}
	\begin{split}
	&= \prod\limits_{i=1}^N(\sum^L_{j=1}U_{jl}(t)\sum_{m=1}^{L} \left[e^{-ih_\text{eff}\delta t}\right]_{m,j}c^{\dagger}_m)|0\rangle\\
	&=\prod\limits_{i=1}^N\sum_{m=1}^{L}\left[e^{-ih_\text{eff}\delta t}U\right]_{m,i}c^{\dagger}_m|0\rangle ,
		\end{split}
	\end{equation}
In a word, $U(t+\delta t) = e^{-ih_\text{eff}\delta t}U(t)$. To preserve $U^{\dagger} U = I$, we can make $QR$ decomposition, $U(t+\delta t) = Q\cdot R$, and reassign $U(t+\delta t)$ as Q

2)
Next, consider the action of quantum jumps 
\begin{equation}
	\begin{split}
	U(t+\delta t)\rangle = M_{\delta t}[e^{-iH_\text{eff}\delta t}|U(t)\rangle],\\
	M_{\delta t}[|U\rangle] = \prod\limits_{i\in P}\dfrac{L_i|U\rangle}{\parallel L_i|U\rangle\parallel},
		\end{split}
	\end{equation}
where $P=\{n|r_n<\gamma\langle L^{\dagger}_nL_n\rangle\delta t\}$, $r_n \in (0,1)$ is a set of independent random variables.
In detail, after the non-Hermitian evolution, we can generate a set of random numbers $r_n$ to  decide whether quantum jump $\hat{L}_n$ will happen. 
Assume $\hat{L}_i=e^{i\pi\hat{n}_{i+1}} \hat{\xi}_{i}^{\dagger}\hat{\xi}_{i}$, $\hat{\xi}^{\dagger}_i=\sum_ka_{ik}\hat{c}_k$,
the probability of the quantum jump $\hat{L}_{i}$ is 
$p_i = \langle\psi|\hat{L}_i^{\dagger}\hat{L}_i|\psi\rangle\gamma\delta t$ = $\langle\psi|\hat{\xi}^{\dagger}_i\hat{\xi}_i|\psi\rangle\gamma\delta t$ = $\parallel \hat{\xi}_i|\psi\rangle\parallel^2\gamma\delta t$
\begin{equation}
	\begin{split}
	\hat{\xi}_i|\psi\rangle = (\sum_ka^*_{ik}\hat{c}_k)\prod_{i=1}^N(\sum_{j=1}^LU_{jl}(t)c^{\dagger}_j)|0\rangle\\
	=\sum_i\langle a|U_i\rangle\bigotimes_{j\neq i}|U_j\rangle,
		\end{split}
	\end{equation}
Because of $U^{\dagger}U = I$, we can get $p_i=\sum_{i}|\langle a|U_{i}\rangle|^{2}$.
To simplify the expression, remember that the state is invariant for the elementary column 
operations, we can pre-orthogonalize, find the first column $i$  which satisfies $\langle a|U_i\rangle\neq 0$, then move the column $i$ into the first column, and transform the left columns as 
\begin{equation}
	\begin{split}
	|U_{j}^{'}\rangle &= |U_j\rangle-\dfrac{\langle a|U_j\rangle}{\langle a|U_1\rangle}|U_1\rangle,
		\end{split}
	\end{equation}
So $\forall j\geq 2$, \, $\langle a|U_{j}^{'} \rangle = 0$, and $\hat{\xi}_{i}|\psi\rangle= \bigotimes_{i\geq 2}|U_{i}^{'}\rangle$. After the action of Lindblad operator, $\hat{L}_{i}|\psi\rangle=e^{i\pi\hat{n}_{i+1}}\hat{\xi}_{i}^{\dagger}\hat{\xi}_{i}|\psi\rangle=e^{i\pi \hat{n}_{i+1}}\left[| a\rangle\bigotimes_{i\geq 2}|U_{i}^{'}\rangle\right]=|e^{i\pi M} a\rangle\bigotimes_{i\geq 2}|e^{i\pi M}U_{i}^{'}\rangle$, in which $M$ is $L \times L$ matrix, with only one nonzero element $M_{i+1,i+1}=1$. After the action of quantum jump $\hat{L}_{i}$, perform the QR decomposition, and reassign $U$ as Q. 

\section{Numerical results}
\label{sec:appendixB}

\begin{figure}[htb]
\centering
\centering
\includegraphics[height=3.0cm,width=8.5cm]{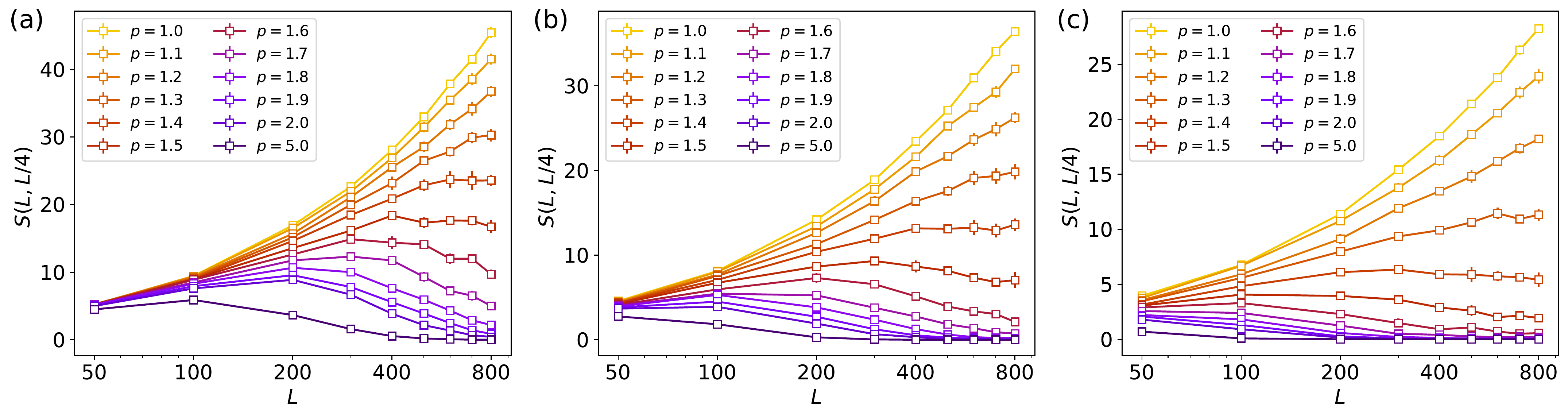}

\caption{Entanglement entropy $S(L, L/4)$. (a) $\gamma=0.05$, (b) $\gamma=0.1$, (c) $\gamma=0.2$. With the decrease of $p$, the hopping range is extended, hence, the skin effect gets weaker and entanglement entropy improves. With the increase of monitoring rate $\gamma$, the skin effect gets stronger and entanglement entropy lowers down.}
\label{suppfig1}
\end{figure}

As shown in Fig.  \ref{suppfig1}, for fixed monitoring rate $\gamma$, with the decrease of $p$, namely hopping range getting extended, entanglement entropy grows. Moreover, the threshold $L_{0}$, in which entanglement entropy begins to decay, also increases. While for  fixed $p$, with the increase of monitored rate $\gamma$, the skin effect gets stronger, thus entanglement entropy lowers down, and the threshold $L_{0}$ also decreases. Worth to mention, for $p=1.0, 1.1, 1.2$, although entanglement entropy always increases with system size $L$, however, in the main text, we have pointed out it's a finite-size effect. To observe the decaying behavior of entanglement entropy for small $p$, the results with larger system sizes $L$ are required, which is beyond our numerical capabilities.

\begin{figure}[htb]
\centering
\centering
\includegraphics[height=6.0cm,width=8.6cm]{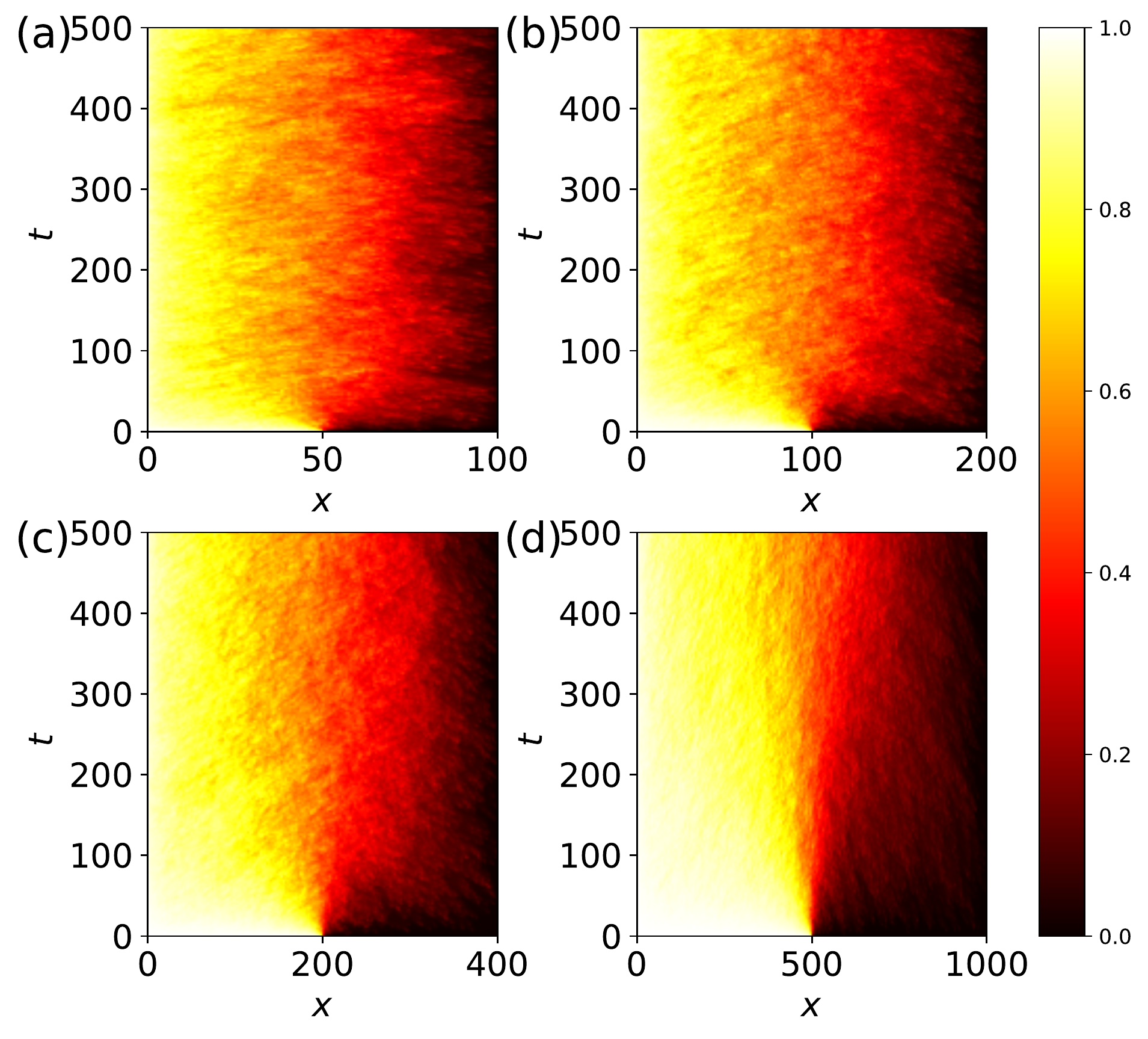}

\caption{Density distribution evolution with time. The initial state is chosen as $|111..000\rangle$. $p=1.1,\gamma=0.2$, with increasing of system size $L$ ((a) $L=100$, (b) $L=200$, (c) $L=400$, (d) $L=1000$), fluctuation areas with respect to system size $L$ seem to narrow down}
\label{suppfig2}
\end{figure}

As Fig. \ref{suppfig2} shows, for $p=1.1$ and $\gamma=0.2$, with the increase of system size $L$, the ratio of fluctuation areas' size to system size $L$ tends to decrease. Therefore, we predict in the thermodynamical limit, the size of fluctuation areas will saturate, and entanglement entropy obeys ``area law". For finite sizes we calculate, even for $L=1000$, the fluctuation areas' size is still comparable with the system size $L$ and always grows with $L$, which explains why entanglement entropy grows with $L$ all along in Fig. \ref{suppfig1} for $p$ close to 1. However, as shown in  Fig. \ref{Delta n}(b), for $p=1.1$ and $\gamma=0.2$, $\Delta n$ grows with $L$, which means the skin effect gets strengthened.

\begin{figure*}[htb]
\centering
\includegraphics[height=7.2cm,width=17.6cm]{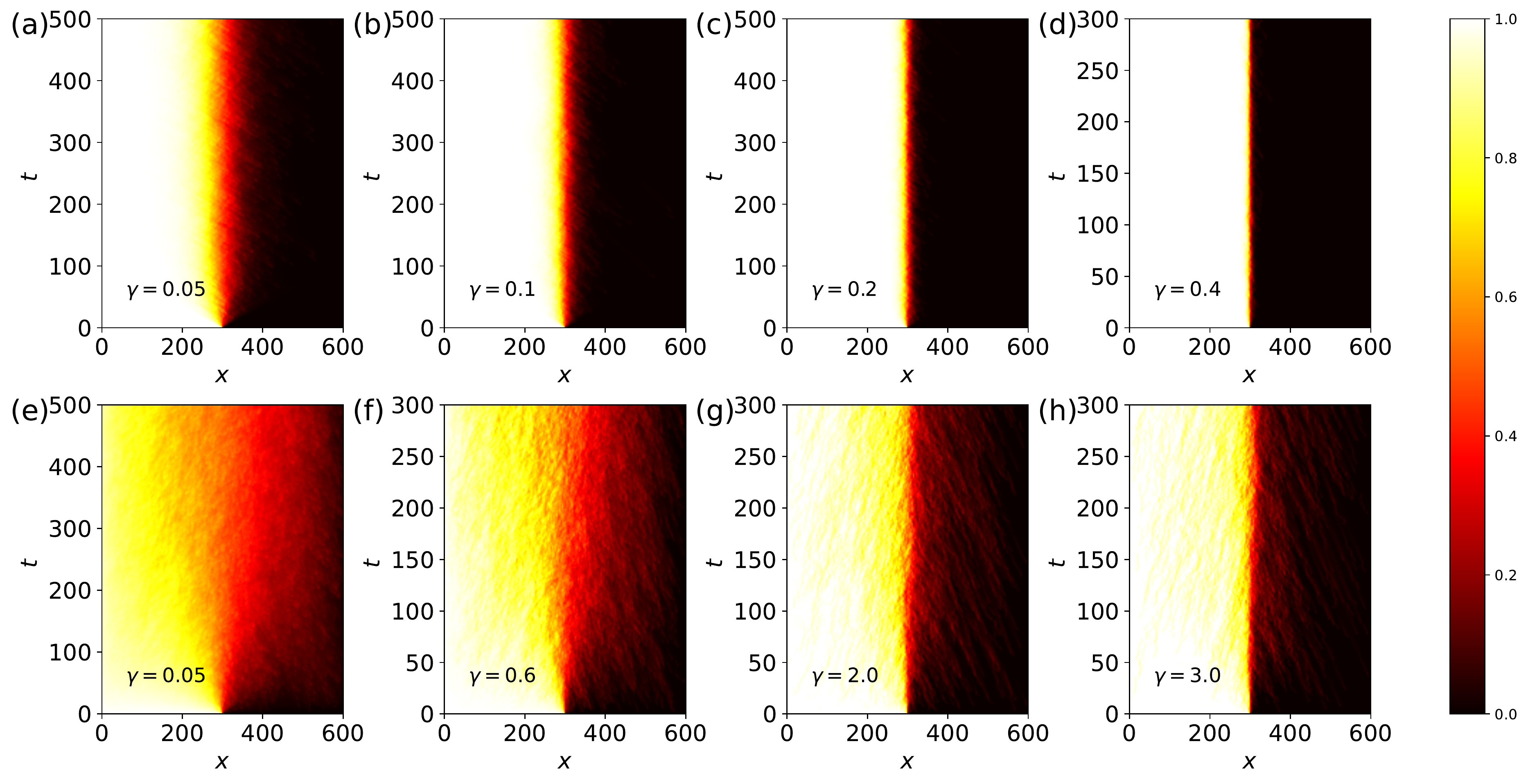}

\caption{Density distribution evolution with time. The initial state is chosen as $|111..000\rangle$. above: $p=2.0$, bottom: $p=1.1$, for $p=2.0$, fluctuation areas are small; for $p=1.1$, longer-range hopping greatly extends fluctuation areas. Both $p=1.1$ and $p=2.0$ show that the skin effect is strengthened with the increase of $\gamma$.}
\label{suppfig3}
\end{figure*}

Comparing the above pictures with the bottom pictures in Fig. \ref{suppfig3}, it's clear that the extended hopping range greatly improves the size of fluctuation areas. For $p=2$, the density distributions are very close to $p=\infty$ case \cite{MISE}, which show the extremely strong skin effect with almost one-half occupied and one-half unoccupied. As for $p=1.1$ and small $\gamma$, when system size $L$ is about $O(10^{2})$, fluctuation areas' size is comparable with $L$. Therefore, entanglement entropy for $p=2.0$ quickly decays to zero with the increase of $L$, while for $p=1.1$ and small $\gamma$, at least for $L\leq 800$ entanglement entropy always grows with $L$ (Fig. \ref{suppfig1}). 

\begin{figure}[H]
\centering
\includegraphics[height=6.6cm,width=8.0cm]{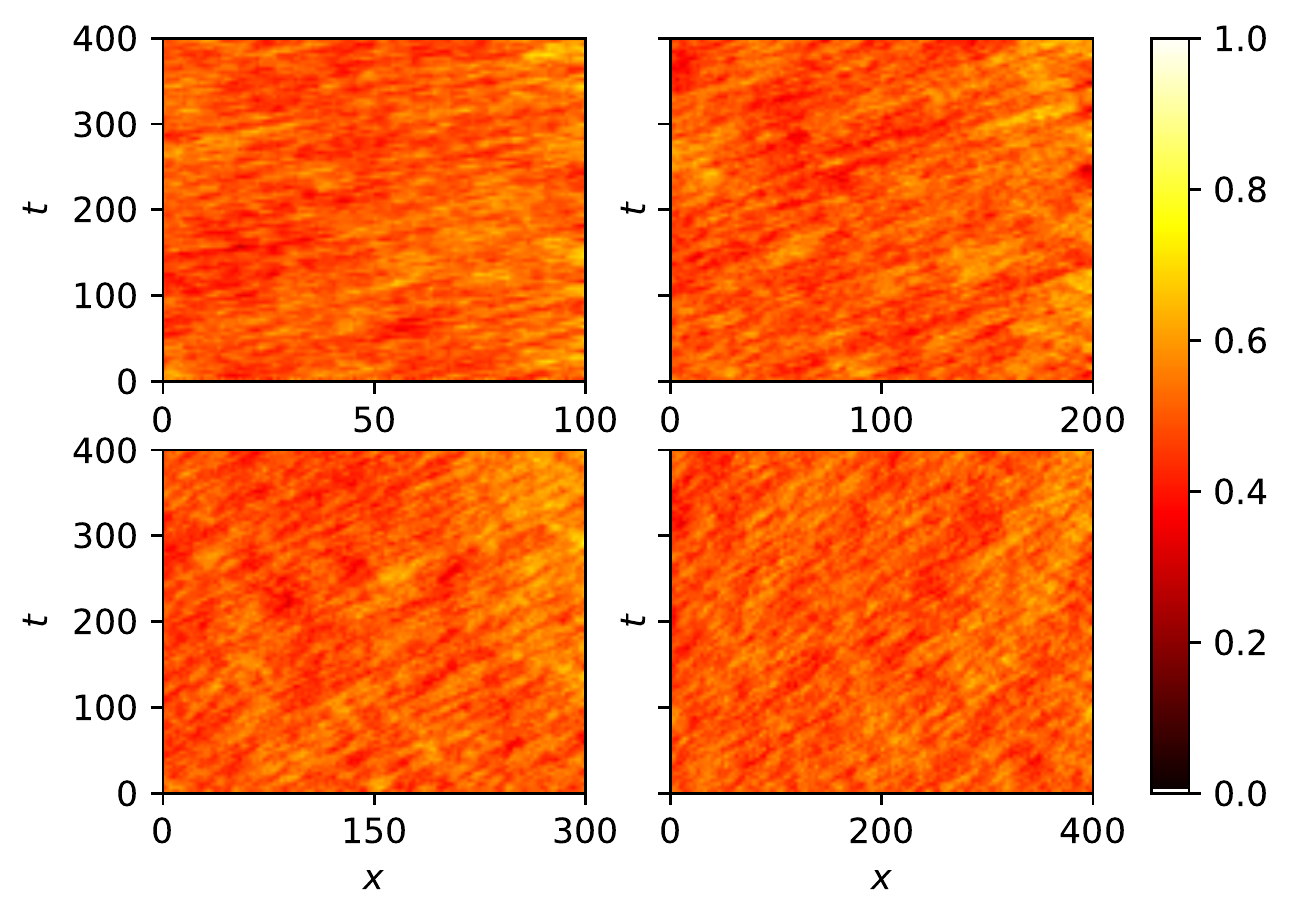}

\caption{Density distribution evolution with time without feedback under OBC. The initial state is chosen as $|1010..010\rangle$. For $p=2.0,\gamma=0.4$, time step $\delta t$=0.05 and various sizes $L$, density distributions are uniform.}
\label{suppfig4}
\end{figure}

For the case without feedback, from Lindblad master equation we know there is no skin effect in steady state, which is demonstrated in Fig. \ref{suppfig4}. For relatively small $\gamma=0.4$, time step $\delta t=0.05$, we can already see the steday state's density  distribution is uniform.

\end{document}